\documentclass[
preprint,
superscriptaddress, prd,tightenlines,showpacs,nofootinbib, eqsecnum,amsfonts,amsmath,amssymb]{revtex4}

\usepackage{bm}
\usepackage{graphicx}
\usepackage{hyperref}
\usepackage{color}

\newcommand{\Z}{\mathbb{Z}}
\newcommand{\N}{\mathbb{N}}

\newcommand{\ud}{\mathrm{d}}
\newcommand{\ab}{^{\alpha\beta}}
\newcommand{\lab}{_{\alpha\beta}}

\allowdisplaybreaks

\newcommand{\ua}{^{\alpha}}
\newcommand{\ub}{^{\beta}}
\newcommand{\la}{_{\alpha}}

\newcommand{\ubar}{{\bar u}}

\newcommand{\ret}{{\text{ret}}}
\newcommand{\s}{{\text{S}}}
\newcommand{\R}{{\text{R}}}
\newcommand{\MU}{{m_1}}
\newcommand{\m}{{m_2}}

\newcommand{\beq}{\begin{equation}}
\newcommand{\eeq}{\end{equation}}

\begin{document}

\title{Post-Newtonian and Numerical Calculations of the Gravitational Self-Force for Circular Orbits in the Schwarzschild Geometry}
\author{Luc Blanchet}\email{blanchet@iap.fr}
\affiliation{$\mathcal{G}\mathbb{R}\varepsilon{\mathbb{C}}\mathcal{O}$, Institut d'Astrophysique de Paris --- UMR 7095 du CNRS, \\ Universit\'e Pierre \& Marie Curie, 98\textsuperscript{bis} boulevard Arago, 75014 Paris, France}
\author{Steven Detweiler}\email{det@phys.ufl.edu}
\affiliation{Institute for Fundamental Theory, Department of Physics, University of Florida, Gainesville, FL 32611-8440, USA}
\author{Alexandre Le Tiec}\email{letiec@iap.fr}
\affiliation{$\mathcal{G}\mathbb{R}\varepsilon{\mathbb{C}}\mathcal{O}$, Institut d'Astrophysique de Paris --- UMR 7095 du CNRS, \\ Universit\'e Pierre \& Marie Curie, 98\textsuperscript{bis} boulevard Arago, 75014 Paris, France}
\author{Bernard F. Whiting}\email{bernard@phys.ufl.edu}
\affiliation{Institute for Fundamental Theory, Department of Physics, University of Florida, Gainesville, FL 32611-8440, USA}

\date{\today}

\begin{abstract}

The problem of a compact binary system whose components move on circular orbits is addressed using two different approximation techniques in general relativity. The post-Newtonian (PN) approximation involves an expansion in powers of $v/c\ll 1$, and is most appropriate for small orbital velocities $v$. The perturbative self-force (SF) analysis requires an extreme mass ratio $m_1/m_2\ll 1$ for the components of the binary. A particular coordinate-invariant observable is determined as a function of the orbital frequency of the system using these two different approximations. The post-Newtonian calculation is pushed up to the third post-Newtonian (3PN) order. It involves the metric generated by two point particles and evaluated at the location of one of the particles. We regularize the divergent self-field of the particle by means of dimensional regularization. We show that the poles $\propto (d-3)^{-1}$ appearing in dimensional regularization at the 3PN order cancel out from the final gauge invariant observable. The 3PN analytical result, through first order in the mass ratio, and the numerical SF calculation are found to agree well. The consistency of this cross cultural comparison confirms the soundness of both approximations in describing compact binary systems. In particular, it provides an independent test of the very different regularization procedures invoked in the two approximation schemes.
\end{abstract}


\maketitle

\section{Introduction}
\label{intro}

\subsection{Motivation}
\label{motiv}

The detection and analysis of the gravitational radiation from black hole binaries by the ground-based LIGO--VIRGO and space-based LISA observatories requires very accurate theoretical predictions, for use as gravitational wave templates \cite{Th300}. There are two main approximation schemes available for performing such calculations in general relativity: (i) The \textit{post-Newtonian} expansion, well suited to describe the inspiralling phase of arbitrary mass ratio compact binaries in the slow motion and weak field regime ($c^{-1} \equiv v / c \ll 1$),\footnote{By a slight abuse of notation we denote by $c^{-1}$ the standard PN estimate, where $c$ is the speed of light. As usual we refer to $n$PN as the order equivalent to terms $\mathcal{O}(c^{-2n})$ in the equations of motion beyond the Newtonian acceleration.} and (ii) the perturbation-based \textit{self-force} approach, which gives an accurate description of extreme mass ratio binaries ($q \equiv m_1/m_2 \ll 1$) even in the strong field regime.

For the moment the post-Newtonian (PN) templates for compact binary inspiral have been developed to 3.5PN order in the phase \cite{BIJ,BFIJ,BDEI04,BDEI05dr} and 3PN order in the amplitude \cite{Kidder,BFIS} (see \cite{Bliving} for a review). These are suitable for the inspiral of two neutron stars in the frequency bandwidth of LIGO and VIRGO detectors. For detection of black hole binaries (with higher masses) the PN templates have to be matched with full numerical simulations for the merger phase and the ringdown of the final black hole. The matching between the PN approximation and numerical relativity has turned out to be very successful \cite{BCP07,Boyle}.

On the other hand, gravitational self-force (SF) analysis \cite{MiSaTa,QuWa,DW03,GW08,Poisson} is expected to provide templates for extreme mass ratio inspirals (EMRIs) anticipated to be present in the LISA frequency bandwidth. SF analysis is a natural extension of first order perturbation theory, and the latter has a long history of comparisons with post-Newtonian analysis \cite{P93a,CFPS93,P93b,TNaka94,P95,TTS96,TSTS96,PS95}. SF analysis, itself, is just now mature enough to present some limited comparisons with PN analysis, but it is not yet ready for template generation.

\begin{figure}
    \includegraphics[width=11cm]{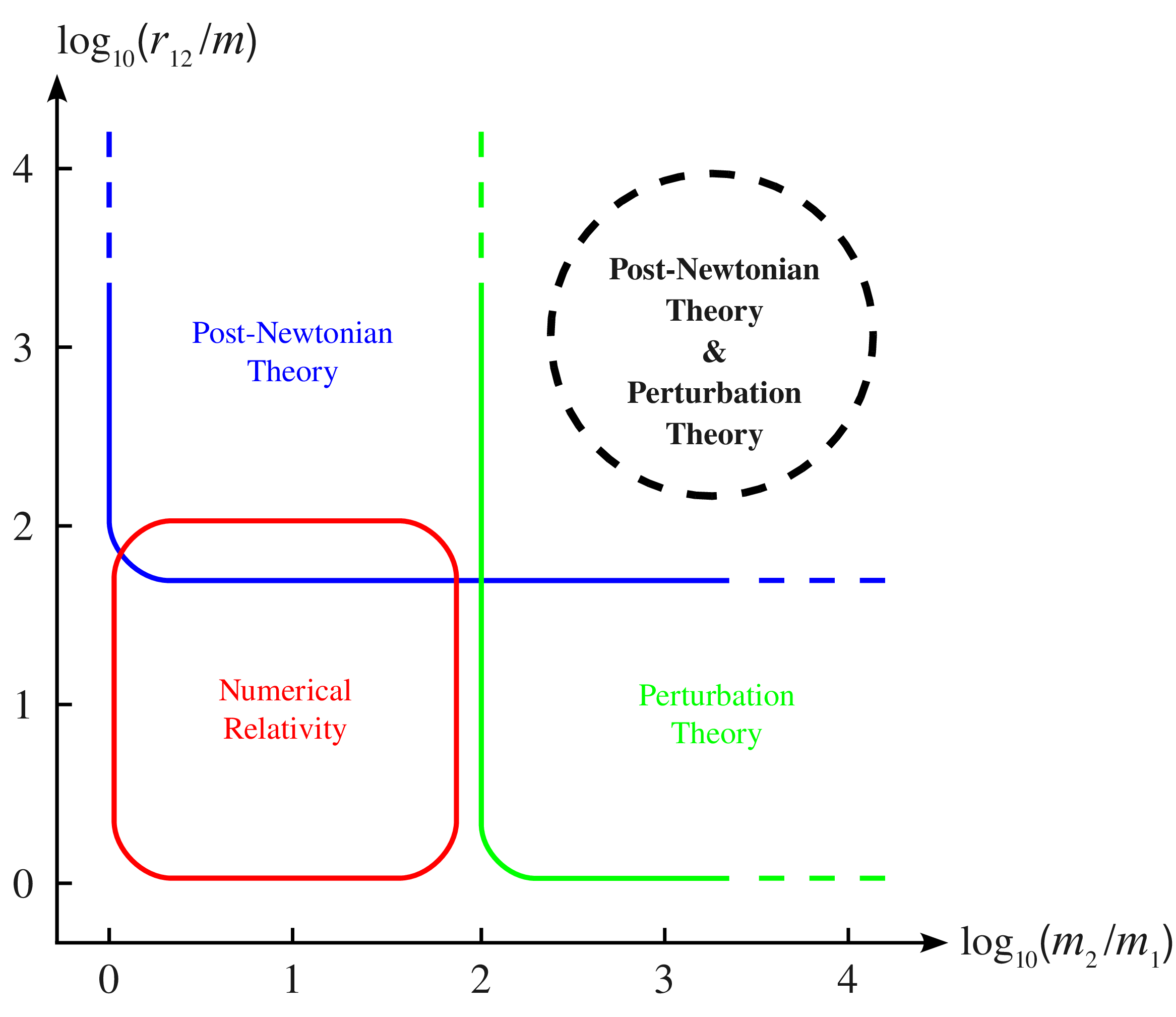}
    \caption{\footnotesize Different analytical approximation schemes and numerical techniques are used to study black hole binaries, depending on the mass ratio $m_1/m_2$ and the orbital velocity $v^2 \sim G m/r_{12}$, where $m = m_1 + m_2$. The post-Newtonian theory and black hole perturbation theory can be compared in the slow motion regime ($v\ll c$ equivalent to $r_{12}\gg G m/c^2$ for circular orbits) of an extreme mass ratio ($m_1 \ll m_2$) binary.}
    \label{methods}
\end{figure}

In this paper we shall compare the PN and SF analyses in their common domain of validity, that of the slow motion weak field regime of an extreme mass ratio binary (see illustration of various methods in Fig.~\ref{methods}). The problem was tackled by Detweiler \cite{Det08}, who computed numerically within the SF a certain gauge invariant quantity, defined by \eqref{ut_defX} below for an extreme mass ratio binary, and compared it with the 2PN prediction extracted from existing PN results \cite{BFP98}. Here we shall go one step further, and extend the comparison up to 3PN order. This will require an improvement in the numerical resolution of the SF calculation in order to distinguish more accurately the 3PN self-force from the self-force at higher PN orders. However, our primary difficulty is that the PN results for the metric have not previously been available at 3PN order, and will have to be carefully derived. We shall demonstrate an excellent agreement between the extreme mass ratio case ($q \ll 1$) of the analytical 3PN result and the numerical SF result.

\subsection{Method}
\label{method}

Let us consider a system of two (non-spinning) compact objects with masses $m_1$ and $m_2$, and moving on slowly inspiralling quasi-circular orbits. In the PN analysis, let $m_1$ and $m_2$ be arbitrary; in the SF analysis, further assume that $m_1 \ll m_2$. We can then call $m_1$ the ``particle'', and $m_2$ the ``black hole''.

Self-force analysis shows that the dissipative parts of the self-force for a circular orbit are the $t$ and $\varphi$ components. These result in a loss of energy and angular momentum from the small mass at the same precise rate as energy and angular momentum are radiated away\cite{Det08}. In addition, earlier perturbative calculations of energy and angular momentum fluxes \cite{P93a,CFPS93,P93b,TNaka94,P95,TTS96,TSTS96,PS95} for this situation show them to be equivalent to the results of the PN analysis in their common domain of validity. Hence, by invoking an argument of energy and angular momentum balance, we know that the PN results also agree with the dissipative parts of the SF in their domain of common validity, and further comparison can reveal nothing new.

For our PN-SF comparison, we shall thus neglect the dissipative, radiation-reaction force responsible for the inspiral, and restrict ourselves to the conservative part of the dynamics. In PN theory this means neglecting the dissipative radiation-reaction force at 2.5PN and 3.5PN orders, and considering only the conservative dynamics at the even-parity 1PN, 2PN and 3PN orders. This clean separation between conservative even-parity and dissipative odd-parity PN terms is correct up to 3.5PN order.\footnote{However, this split merges at 4PN order, since at that approximation arises a contribution of the radiation-reaction force, which originates from gravitational wave tails propagating to infinity \cite{BD88}.} In SF theory there is also a clean split between the dissipative and conservative parts of the self-force. This split is particularly transparent for a quasi-circular orbit, where the $r$ component is the only non-vanishing component of the conservative self-force.

Henceforth, the orbits of both masses are assumed to be and to remain circular, because we are ignoring the dissipative radiation-reaction effects. For our comparison we require two physical quantities which are precisely defined in the context of each of our approximation schemes. The orbital frequency $\Omega$ of the circular orbit as measured by a distant observer is one such quantity. The second requires further explanation.

With circular orbits and no dissipation, the geometry has a helical Killing vector field $k\ua$. A Killing vector is only defined up to an overall constant factor. In our case $k^\alpha$ extends out to a large distance where the geometry is essentially flat. There $k\ua\partial\la = \partial_t + \Omega\,\partial_\varphi$ in any natural coordinate system which respects the helical symmetry \cite{SBD08}. We let this equality define the overall constant factor, thereby specifying the Killing vector field uniquely.

An observer moving with the particle $m_1$, while orbiting the black hole $m_2$, would detect no change in the local geometry. Thus the four-velocity $u_1^\alpha$ of the particle is tangent to the Killing vector $k^\alpha$ evaluated at the location of the particle, which we denote by $k_1\ua$. A second physical quantity is then defined as the constant of proportionality, call it $u_1^T$, between these two vectors, namely
\begin{equation}
   u_1\ua = u_1^T \,k_1\ua\,.
\label{ut_def}
\end{equation}
The four-velocity of the particle is normalized so that ${(g\lab)}_1 u_1\ua u_1\ub = -1$; ${(g\lab)}_1$ is the \textit{regularized} metric at the particle's location, whereas the metric itself is formally singular at the particle $m_1$ in both PN and SF approaches. The gauge invariant quantity $u_1^T$ is thus given by:
\begin{equation}
   u_1^T = \left( - {(g\lab)}_1 u_1\ua k_1\ub \right)^{-1} = \left( - {(g\lab)}_1 k_1\ua k_1\ub \right)^{-1/2} .
\label{u1T}
\end{equation}
It is important to note that this quantity is precisely defined in both PN and SF frameworks, and it does not depend upon the choice of coordinates or upon the choice of perturbative gauge; however, it very definitely depends upon using a valid method of regularization. Furthermore, for any coordinate system $u_1^T$ has a pleasant physical interpretation as being the rate of change of time at a large distance, with respect to the proper time on the particle $m_1$, and it could in principle be measured by a redshift experiment as described in \cite{Det08}.

If we happen to choose a convenient coordinate system where $k\ua\partial\la = \partial_t + \Omega\,\partial_\varphi$ everywhere, then in particular $k_1^t=1$, and thus $u_1^T\equiv u_1^t$, the $t$ component of the four velocity of $m_1$. The Killing vector on the particle is then $k_1\ua=u_1\ua/u_1^t$, and simply reduces to the particle's ordinary post-Newtonian coordinate velocity $v_1\ua/c$. In such a coordinate system, the description of the invariant quantity we are thus considering is
\begin{equation}\label{ut_defX}
    u_1^T \equiv u_1^t = \biggl( - {(g\lab)}_1 \frac{v_1\ua v_1\ub}{c^2} \biggr)^{-1/2} .
\end{equation}
In the PN calculation we shall evaluate $u_1^T$ using a particular harmonic coordinate system. We shall make no restriction on the mass ratio $q=m_1/m_2$, but shall eventually compute the small mass ratio limit $q \ll 1$ for comparison with the SF result.

The regularized metric ${(g\lab)}_1$ is defined with very different prescriptions in the SF and PN approaches. Both analyses require subtle treatment of singular fields at the location of the masses. Subtracting away the infinite part of a field while carefully preserving the part which is desired is always a delicate task. Our comparison will rely on the principle of the \textit{physical equivalence} of the regularized SF and PN metrics, at least in the vicinity of the particle, i.e. that they are isometric --- they differ by a coordinate transformation. In fact the cross cultural comparison of the invariant $u_1^T$ is a test of the isometry of the two regularized metrics and is, thus, a test of the two independent (and very different) regularization procedures in use.

In the SF prescription, the regularized metric reads
\begin{equation}\label{regSF}
    g\lab^\mathrm{SF}(x)=\bar{g}\lab(x)+h\lab^\mathrm{R}(x)\,,
\end{equation}
where $\bar{g}\lab$ denotes the background Schwarzschild metric of the black hole, and where the ``Regular'' perturbation $h\lab^\mathrm{R}$ is smooth in a neighborhood of the particle, and follows from the Detweiler-Whiting prescription \cite{DW03} for removing the infinite part of the field, as described below in Sec.~\ref{modesum}. In particular the metric \eqref{regSF} is regular at the particle's position $y_1\ua$, and we simply have
\begin{equation}\label{regSF1}
    {(g\lab^\mathrm{SF})}_1=g\lab^\mathrm{SF}(y_1)\,.
\end{equation}
In the perturbative SF analysis we are only working through first order in $q=m_1/m_2$, and at that level of approximation $h^\R\lab = \mathcal{O}(q)$. Then $u_1^T$ can be computed accurately to the same perturbative order and compares well with the post-Newtonian result to 2PN order \cite{Det08}. The regularized 2PN metric is known \cite{BFP98}, and therefore the comparison is straightforward.

In the present paper we shall obtain the 3PN regularized metric which will be the core of our calculation, and will be partly based on existing computations of the equations of motion at 3PN order using Hadamard \cite{BFeom} and dimensional \cite{BDE04} regularizations. Using an iterative PN procedure, one first considers the post-Newtonian metric $g\lab^\mathrm{PN}(\mathbf{x},t)$ at any field point outside the particle, in a coordinate system $x\ua=\{ct,x^i\}$. That metric is generated by the two particles, and includes both regular and singular contributions around each particle. Then we compute the PN regularized metric at the location of the particle by taking the limit when $\mathbf{x}\rightarrow\mathbf{y}_1(t)$, where $\mathbf{y}_1(t)$ is the particle's trajectory. In 3 spatial dimensions, that limit is singular. In order to treat the infinite part of the field, we extend the computation in $d$ spatial dimensions, following the prescription of \textit{dimensional regularization}, which is based on an analytic continuation (AC) in the dimension $d$ viewed as a complex number. Considering the analytic continuation in a neighborhood of $\varepsilon\equiv d-3 \rightarrow 0$, we define
\begin{align}\label{gPNdr}
    {(g\lab^\mathrm{PN})}_1 = \mathop{\mathrm{AC}}_{\varepsilon\rightarrow 0}\,\Bigl[\lim_{\mathbf{x}\rightarrow\mathbf{y}_1} g\lab^\mathrm{PN}(\mathbf{x},t)\Bigr]\,.
\end{align}
The limit $\varepsilon\rightarrow 0$ does not exist in general due to the presence of poles $\propto\varepsilon^{-1}$ occurring at 3PN order; we compute the singular Laurent expansion when $\varepsilon\rightarrow 0$, and we shall see that the poles disappear from the final gauge invariant results. Previous work on equations of motion and radiation field of compact binaries has shown that dimensional regularization is a powerful regularization method in a PN context. In particular this regularization is free of the ambiguities plaguing the Hadamard regularization at the third post-Newtonian order \cite{DJSdim,BDE04,BDEI04,BDEI05dr}. 

The plan of this paper is as follows: Sec.~\ref{overview} is devoted to an overview of the SF formalism. The circular geodesics of the perturbed Schwarzschild geometry are described in Sec.~\ref{geod}, where we also give an explicitly gauge invariant relationship between  $\Omega$ and $u_1^T$ for the particle $m_1$. We use the mode-sum regularization procedure of Barack and Ori \cite{BO00,BMNOS02} to perform the delicate subtraction of the singular field $h^\s\lab$ from the retarded metric perturbation $h^\ret\lab$. We give a brief description of our application of this process in Sec.~\ref{modesum}. In Sec.~\ref{pnsf} we describe some of the details of the numerical analysis which yields our value for $u_1^T$ as a function of $\Omega$, and provide a brief discussion of the numerical determination of the 3PN effect on $u_1^T$. Most of the details concerning the 3PN calculation are presented in Sec.~\ref{PN}. We focus mainly on the issues regarding our implementation of the dimensional self-field regularization which is described in Secs.~\ref{metricd} and \ref{reg}. The post-Newtonian results are presented in Sec.~\ref{PNres}. We give the fully-fledged regularized 3PN metric in Sec.~\ref{PNmetric}, and present our final result for $u_1^T$ in Sec.~\ref{uT}. We finally investigate the small mass ratio limit $q\ll 1$ of the post-Newtonian result, and compare with the self-force calculation in Sec.~\ref{comp}. Two appendices provide further details on the PN calculation: An alternative derivation using the Hadamard regularization is discussed in Appendix~\ref{HR}, and the choice of the center-of-mass frame and the reduction to quasi-circular orbits at 3PN order within dimensional regularization are investivated in Appendix~\ref{circ}.

\section{Self-force overview}
\label{overview}

Previously we described the truly coordinate and perturbative-gauge independent properties of $\Omega$ and $u_1^T$. In this section we use Schwarzschild coordinates, and we refer to ``gauge invariance'' as a property which holds within the restricted class of gauges for which $k^\alpha \partial_\alpha = \partial_t + \Omega\,\partial_\varphi$ is a helical Killing vector. In all other respects, the gauge choice is arbitrary. With this assumption, no generality is lost, and a great deal of simplicity is gained.

The regularized metric perturbation $h^\R\lab = h^\ret\lab - h^\s\lab$ is the difference between the retarded metric perturbation $h^\ret\lab$ and the singular field $h^\s\lab$. A Hadamard expansion of Green's functions in curved spacetime provides an expansion for $h^\s\lab$ \cite{DW03}. In a neighborhood of the particle with a special, locally-inertial coordinate system, $h^\s\lab$ appears as the $m_1/r$ part\footnote{In all of Sec.~\ref{overview} we set $G=c=1$.} of the particle's Schwarzschild metric along with its tidal distortion caused by the background geometry of the large black hole. Details of the expansion are given in Sec.~6.1 of \cite{Det05}. The special locally inertial coordinates for a circular geodesic in the Schwarzschild metric are given as functions of the Schwarzschild coordinates in Appendix~B of \cite{DMW03}.

\subsection{Circular geodesics of the perturbed Schwarzschild geometry}
\label{geod}

The effect of the gravitational self-force is most easily described as having $m_1$ move along a geodesic of the regularized metric $\bar g\lab + h^\R\lab$. We are interested in circular orbits and let $u\ua$ be the four-velocity of $m_1$.\footnote{Since we are clearly interested in the motion of the small particle $m_1$, we remove the index $1$ from $u_1\ua$.} This differs from the four-velocity $\ubar\ua$ of a geodesic of the straight Schwarzschild geometry at the same radial coordinate $r$ by an amount of $\mathcal{O}(q)$. Recall that we are describing  perturbation analysis with $q\ll1$, therefore $h^\R\lab = \mathcal{O}(q)$, and all equations in this section necessarily hold only through first order in $q$.

It is straightforward to determine the components of the geodesic equation for the metric $\bar g\lab + h^\R\lab$ \cite{Det08}, and then to find the components of the four-velocity $u\ua$ of $\MU$ when it is in a circular orbit at Schwarzschild radius $r$. We reiterate that the four-velocity is to be normalized with respect to $\bar g\lab + h^\R\lab$ rather than $\bar g\lab$, and that $h^\R\lab$ is assumed to respect the symmetry of the helical Killing vector. In this case we have
\begin{subequations}
\begin{eqnarray}
  (u^t)^2 &=&
    \frac{{r}}{{r}-3{\m} }  \Big[ 1+\ubar^\alpha \ubar^\beta h^\R\lab
       - \frac{{r}}{2}\ubar^\alpha \ubar^\beta\partial_r  h^\R\lab \Big] \, ,
\label{uTinitial}
\\
  (u^\varphi)^2 &=&
      \frac{{r}-2{\m}}{{r}({r}-3{\m} )}
         \biggl[ \frac{{\m} (1+\ubar^\alpha \ubar^\beta h^\R\lab)}{{r}({r}-2{\m})}
     - \frac{1}{2} \ubar^\alpha \ubar^\beta\partial_r h^\R\lab \biggr]\,.
\label{uPeqn}
\end{eqnarray}\end{subequations}
A consequence of these relations is that the orbital frequency of $m_1$ in a circular orbit about a perturbed Schwarzschild black hole of mass $\m$ is, through first order in the perturbation, given by
\begin{equation}
  \Omega^2 = \biggl(\frac{u^\varphi}{u^t}\biggr)^2
     = \frac{{\m} }{{r}^3}
        - \frac{{r}-3{\m} }{2 {r}^2} \,\ubar\ua \ubar\ub \partial_r h^\R\lab\,.
\label{Omega2}
\end{equation}
The angular frequency $\Omega$ is a physical observable and is independent of the gauge choice. However the perturbed Schwarzschild metric does not have spherical symmetry, and the radius of the orbit $r$ is not an observable and does depend upon the gauge choice. That is to say, an infinitesimal coordinate transformation of $\mathcal{O}(q)$ might change $\ubar\ua \ubar\ub \partial_r h^\R\lab$. But if it does, then it will also change the radius $r$ of the orbit in just such a way that $\Omega^2$ as determined from \eqref{Omega2} remains unchanged. Both $u^t\equiv u^T$ and  $u^\varphi \equiv \Omega \, u^T$ are gauge invariant as well.

Our principle interest is in the relationship between $\Omega$ and $u^T$, which we now establish directly using \eqref{uTinitial} and \eqref{Omega2}, writing all equations through first order.  First, we can rearrange \eqref{Omega2} to get:
\beq
\biggl(\frac{{\m} }{{r}}\biggr)^3= (\m \Omega)^2+\biggl(\frac{{\m} }{{r}}\biggr)^2\biggl(1-{3\m\over r}\biggr) \biggl( {r\over 2}\,\ubar\ua \ubar\ub \partial_r h^\R\lab \biggr) \, .
\eeq
Next, we take the cube root of both sides and expand on the right-hand-side (RHS) to obtain:
\beq
\frac{{\m} }{{r}}= (\m \Omega)^{2/3}+{1\over 3}\biggl(\frac{{\m} }{{r}}{1\over (\m \Omega)^{2/3}}\biggr)^2\biggl(1-{3\m\over r}\biggr)\biggl( {r\over 2}\,\ubar\ua \ubar\ub \partial_r h^\R\lab \biggr) \, .
\label{m2overr}
\eeq
The second term on the RHS of \eqref{m2overr} is already first order in $q$.  Thus, in the first two bracketed expressions in this second term, we can replace $\m/r$ by the leading approximation to $\m/r$ from just the first term on the RHS of \eqref{m2overr}, giving:
\beq
\frac{{\m} }{{r}}= (\m \Omega)^{2/3}+{1\over 3}\biggl(1-{3(\m \Omega)^{2/3}}\biggr) \biggl( {r\over 2}\,\ubar\ua \ubar\ub \partial_r h^\R\lab \biggr) \, .
\label{m2overragain}
\eeq
Following \cite{Det08}, we next introduce the gauge invariant measure of the orbital radius
\begin{equation}
  R_\Omega \equiv \left( \frac{m_2}{\Omega^2} \right)^{1/3} \Longrightarrow (\m \Omega)^{2/3}={\m\over R_{\Omega}} \, .
\label{ROmega}
\end{equation}
Now we use this in its second form and substitute back into \eqref{m2overragain}: 
\beq
\frac{{\m} }{{r}}= {\m\over  R_{\Omega}}+{1\over 3} \biggl(1-{3\m\over  R_{\Omega}}\biggr) \biggl( {r\over 2}\,\ubar\ua \ubar\ub \partial_r h^\R\lab \biggr) \, .
\eeq
Multiplying overall by $-3$ and adding 1 to both sides before dividing through, we find:
\beq
{1\over 1-3\m/r} \biggl(1-{r\over 2}\,\ubar\ua \ubar\ub \partial_r h^\R\lab \biggr) = {1\over 1-3\m/R_{\Omega}}\,.
\eeq
This is exactly what we need in \eqref{uTinitial} in order to establish a first order, gauge invariant, algebraic relationship between $u^T$ (to which $u^t$ evaluates in our gauge) and $R_\Omega$ (or equivalently $\Omega$), namely:
\begin{equation}
  (u^T)^2 =
    \left( 1 - \frac{3\m}{R_\Omega} \right)^{-1} \left( 1 + \ubar^\alpha \ubar^\beta h^\R\lab \right) .
\label{uTeqn}
\end{equation}
The lowest order term on the RHS is identical to what is obtained for a circular geodesic of the unperturbed Schwarzschild metric. Indeed, recall that the Schwarzschild part of $u^T$ is known exactly as $u^T_\mathrm{Schw} = \left( 1 - 3 m_2/R_\Omega \right)^{-1/2}$. Thus, if we write
\beq
u^T \equiv u^T_{\mathrm{Schw}}+q\, u^T_{\mathrm{SF}} + \mathcal{O}(q^2) \, ,
\eeq
the first order term in \eqref{uTeqn} gives:
\beq
  q \, u^T_{\mathrm{SF}} = \frac12 \left(1-\frac{3m_2}{R_\Omega}\right)^{-1/2}\!\! \ubar\ua \ubar\ub h^\R\lab \,,
\label{SF}
\eeq
which is $\mathcal{O}(q)$, and contains the effect of the ``gravitational self-force'' on the relationship between $u^T$ and $\Omega$, even though it bears little resemblance to a force. We shall henceforth focus our attention on the calculation of the combination $\ubar^\alpha \ubar^\beta h^\R\lab$.

\subsection{Mode sum regularization}
\label{modesum}

Both the retarded metric perturbation $h^\ret\lab$ and the singular field $h^\s\lab$ are singular at $\MU$. However, we actually determine $h^\ret\lab$ by using the inherent symmetries of the problem to separate variables and to decompose the components of $h^\ret\lab$ in terms of tensor spherical harmonics. Each $\ell,m$ component $h^{\ret\,(\ell,m)}\lab$ is then finite and determined using a standard numerical differential equation solver. Only the sum over modes diverges.

For our problem, we treat the divergence of the singular field $h^\s\lab$ in a related manner. The singular behavior is represented in the known expansion of $h^\s\lab$ about the particle, and is also amenable to a decomposition in terms of spherical harmonics. This procedure is stylistically quite similar to the expansion of the Coulomb field of a point charge, displaced from the origin, in terms of spherical harmonics centered on the origin; this results in the coefficients being proportional to either $1/r^{\ell+1}$ or  $r^\ell$, depending upon whether the field point is inside or outside the charge. In SF analysis, the spherical harmonic coefficients determine the \textit{regularization parameters} of $h^\s\lab$.

Following the original prescription of Barack and Ori \cite{BO00,Bar01} and extending it as in \cite{DMW03}, we first perform the sum over $m$ for the retarded field at the particle
\begin{equation}
 \ubar^\alpha \ubar^\beta h^{\ret\,(\ell)}\lab
    \equiv \sum_{m=-\ell}^{\ell} \ubar^\alpha \ubar^\beta h^{\ret\,(\ell,m)}\lab\,.
\end{equation}
Then we use the recognition that the decomposition of the singular field is of the form
\begin{eqnarray}
  \ubar^\alpha \ubar^\beta h^{\s\,(\ell)}\lab &=&
        B + \frac{C}{\ell+1/2} + \frac{D}{(2\ell-1)(2\ell+3)}
  + \frac{E_1}{(2\ell-3)(2\ell-1)(2\ell+3)(2\ell+5)}
  \nonumber\\&+& \mathcal{O}(\ell^{-6})\,,
\label{FABDE}
\end{eqnarray}
where $B$, $C$, $D$, $E_1$ (and the subsequent $E_2$, $E_3$, etc) are regularization parameters. The particular $\ell$ dependence of the coefficients accompanying the parameters $D$ and $E_n$ is related to the expansion of $(1-\cos\theta)^{n+1/2}$ in terms of Legendre polynomials $P_\ell(\cos\theta)$; details are derived and described in Appendix D of \cite{DMW03}.

The regular field at the particle is finally given by
\beq
  \ubar\ua \ubar\ub h^\R\lab  = \sum_{\ell} \bigl(\ubar\ua \ubar\ub h^{\ret\,(\ell)}\lab - \ubar\ua \ubar\ub h^{\s\,(\ell)}\lab\bigr)\,,
\label{lsum}
\eeq
and the sum is guaranteed to be convergent as long as $B$ and $C$ are known. In practice, the regularization parameters are difficult to determine. For our problem it is known analytically that $C=0$ and
\begin{equation}
  B = {2m_1\over r}\bigg[\frac{r-3\m}{r-2\m}\bigg]^{1/2}
  \!\!{}_2F_1\left(\frac{1}{2},\frac{1}{2},1,\frac{\m}{r-2\m}\right),
\label{coeffB}
\end{equation}
where ${}_2F_1$ is a hypergeometric function, and $r$ is the Schwarzschild radial coordinate of the circular orbit. This knowledge of $B$ and $C$, but not $D$, implies that the sum in \eqref{lsum} converges as $1/\ell$. To increase the rate of convergence, we augment our knowledge of $B$ and $C$ by numerically determining further regularization parameters \cite{DMW03}: We use the fact that the behavior of $\ubar^\alpha \ubar^\beta h^{\ret(\ell)}\lab$, evaluated at the particle, must match $\ubar^\alpha \ubar^\beta h^{\s\,(\ell)}\lab$ as given in \eqref{FABDE} for large $\ell$. This allows us to fit the numerical data to determine the additional regularization parameters $D$ and $E_n$ up to, say, $E_3$. Knowledge of these additional parameters results in a sum which converges as $1/\ell^9$. In our numerical work we typically fit for three or four extra parameters. We calculate up to $\ell=40$, fit in the range $\ell=13-40$, and then sum to $\ell \rightarrow +\infty$, with errors at the full level of our calculational precision.

In earlier work \cite{Det08,Sago:2009zz,SBD08} the accuracy of the numerical integration used was adequate for the purposes then at hand. For the comparisons presented here it became obvious that we should investigate pushing our integration procedure to enable us to obtain the highest precision practicable.  By adjusting the effective step size as $\ell$ changed,\footnote{We used an integration procedure that contained an adjustable parameter, $\epsilon$, which controlled the precision of the numerical result.  We chose $\epsilon$ small enough so that further reduction would not cause relative changes in the result larger than $10^{-15}$.} we found that it was possible to achieve this without encountering any other numerical difficulties (such as an unreasonable accumulation of round-off error). Subsequent fitting, to obtain the numerical determination of the higher order regularization parameters $D$, $\cdots$, $E_3$ as described above, allowed us to reduce residuals to the level of the computational precision which had controlled our integration procedure. Monte Carlo calculations based on these residuals gave us systematic estimates of the errors to associate with our fit parameters. Using these, we find relative errors of order $10^{-13}$ in $u_{\mathrm{SF}}^T$ (the loss in precision being due to the regularization). The corresponding results are presented in Table \ref{data}.
%
\begin{table*}[h]
\begin{center}
\begin{tabular}{c c c}
\hline\hline
$R_\Omega/m_2$ & $\ubar\ua \ubar\ub h^\R\lab/q$ & $u^T_{\text{SF}}$ \\
\hline
$200$ & $\!\!\!-0.0100252390238679 $ & $\quad -0.00505064245513028$\\
$220$ & $-0.00911174844278219$ & $\quad -0.00458725834137915$\\
$240$ & $-0.00835083080996084$ & $\quad -0.00420175898117037$\\
$260$ & $-0.00770720725494635$ & $\quad -0.00387603022007156$\\
$280$ & $-0.00715569723937482$ & $\quad -0.00359717107497568$\\
$300$ & $-0.00667784659538770$ & $\quad -0.00335574417643231$\\
$320$ & $-0.00625982212277844$ & $\quad -0.00314468649077390$\\
$340$ & $-0.00589105041112645$ & $\quad -0.00295860680303681$\\
$360$ & $-0.00556331104384481$ & $\quad -0.00279331869895365$\\
$380$ & $-0.00527011654983391$ & $\quad -0.00264552181684313$\\
$400$ & $-0.00500627861027562$ & $\quad -0.00251257921031088$\\
$420$ & $-0.00476759835869862$ & $\quad -0.00239235862943596$\\
$440$ & $-0.00455064124356486$ & $\quad -0.00228311728867935$\\
$460$ & $-0.00435257068802445$ & $\quad -0.00218341682793388$\\
$480$ & $-0.00417102337921533$ & $\quad -0.00209205962311231$\\
$500$ & $-0.00400401451882955$ & $\quad -0.00200804044413982$\\
\hline\hline
\end{tabular}
\caption{Summary of the gravitational self-force effects for a variety of radii $R_\Omega$. Approximately 13 digits are believed to be accurate.}
\label{data}
\end{center}
\end{table*}

After the regularization procedure is complete we have in hand $\ubar^\alpha \ubar^\beta h^\R\lab$, and hence  $u_\text{SF}^T$, for an orbit at a given radius $R_{\Omega}$. At this point, we have solved our self-force problem --- we have found the effect, $u_\text{SF}^T$, of the self-force on $u^T$ for a specific $\Omega$.

\subsection{Post-Newtonian fit of $u_{\mathrm{SF}}^T$}
\label{pnsf}

The improved quality of the data in Table \ref{data} fed directly into the next stage, that of fitting $u_{\mathrm{SF}}^T$ as a function of $\Omega$ (or $R_{\Omega}$) to determine the higher order post-Newtonian coefficients. In order to proceed to our post-Newtonian fit of $u_{\mathrm{SF}}^T$, we introduce a special notation for the convenient gauge invariant PN parameter defined in \eqref{ROmega}, which is $\mathcal{O}(c^{-2})$ and reads
\begin{equation}
  y \equiv \left(m_2 \Omega \right)^{2/3} = \frac{m_2}{R_\Omega} \, .
\label{yPN}
\end{equation}
The post-Newtonian expansion of the self-force effect given in \eqref{SF} was determined explicitly up to 2PN order in \cite{Det08}, and found to be
\begin{equation}\label{utSF2PN}
    u^T_\mathrm{SF} = - y - 2 y^2 - 5 y^3 + {\cal C}^\mathrm{SF}_\text{3PN}\, y^4 + \mathcal{O}(y^5) \,,
\end{equation}
where ${\cal C}^\mathrm{SF}_\text{3PN}$ represents the 3PN coefficient (unknown at the time of \cite{Det08}), and higher-order 4PN terms are neglected. The author of Ref.~\cite{Det08} also performed a numerical fit of $u^T_{\mathrm{SF}}$ to the polynomial \eqref{utSF2PN} in order to determine the numerical value of ${\cal C}^\mathrm{SF}_\text{3PN}$. He expected that the post-Newtonian derivation of this coefficient would be unavailable for some time, and the numerical fit was done in a cursory fashion using a range in $R_\Omega$ from $20 \, m_2$ to $50 \, m_2$, not generally optimal for PN comparison. It is now time to improve upon that early analysis.

The process of fitting terms in the $\ell$-sum for the regularization parameters is relatively easy. Convergence in the $\ell$-sum increases by two orders with each additional regularization parameter, and is very rapid. By contrast, the low order PN series for $u_{\mathrm{SF}}^T$ is effectively a power series in $1/R_{\Omega}$ and is relatively slowly convergent. Moreover, if we tried to fit higher terms, we could rapidly encounter the situation where, for some large $R_\Omega$, contributions would be below our error estimates, while for smaller $R_\Omega$, the same contributions would still be significant. This situation complicated both the choice of the range of $R_\Omega$ over which we could effectively fit, and the choice of the number of additional PN coefficients we should use to improve the characterization of our available data, consistent with the error estimates we had previously established. For this paper, we settled on a compromise, namely, we used values of $R_\Omega$ generally in the range $200\,m_2$ to $500\,m_2$ and, surprisingly, somewhere between 3 and 6 additional PN coefficients. Following these procedures, our numerical determination of the 3PN coefficient in the self-force effect upon $u^T$ was found to be
\begin{equation}\label{CSF}
\mathcal{C}_\text{3PN}^\text{SF} = - 27.677\pm 0.005\, .
\end{equation}
If we do not use a sufficient number of additional coefficients, our fitting procedure will compensate by systematically attempting to approximate the missing terms  by adjusting the fitting parameters we do use. This effect, which determines the error term in \eqref{CSF}, turns out to dominate the random error from our numerically determined data points. We can estimate this effect by our choice of the range of data and the number of coefficients used in the fitting process. This proves to be adequate for our comparison while also suggesting that further work is warranted. In particular, we shall show in separate work \cite{BDLW09b} that the PN expansion involves in higher orders some logarithmic terms, and that the prior knowledge of the coefficients of the logarithms appearing at 4PN and 5PN orders (computed in \cite{BDLW09b} from PN theory) will improve very much the accuracy of the PN fit to the SF result.\footnote{Accordingly, the $\mathcal{O}(y^5)$ symbol for remainders in Eq.~\eqref{utSF2PN} and similar equations below should rather be understood as the Landau $o(y^4)$ symbol.} This study is beyond our current scope, but will be extensively reported in \cite{BDLW09b}.

\section{Post-Newtonian calculation}
\label{PN}

In this section, our aim is to compute the 3PN regularized metric \eqref{gPNdr} by direct post-Newtonian iteration of the Einstein field equations in the case of singular point mass sources. Previous work on the 3PN equations of motion and radiation field of point particles \cite{DJSdim,BDE04,BDEI04,BDEI05dr} has shown that the appropriate regularization to remove the infinite self-field of point particles in this context is dimensional regularization \cite{tHooft,Bollini}.

In the dimensional regularization (DR) scheme, we look for the solution of the Einstein field equations in $D=d+1$ space-time dimensions, with a matter source made of point particles. We treat the space dimension as an arbitrary complex number, $d\in\mathbb{C}$, and interpret any intermediate formula in the PN iteration of those equations by analytic continuation in $d$. Then we analytically continue $d$ down to the value of interest (namely 3), posing
\begin{equation}\label{deps}
d \equiv 3 + \varepsilon\,.
\end{equation}
In most of the calculations we neglect terms of order $\varepsilon$ or higher, i.e. we retain the finite part and the eventual poles.

Defining the gravitational field variable $h\ab\equiv\sqrt{-g}\, g\ab - \eta\ab$,\,\footnote{Here $g\ab$ is the contravariant metric, inverse of the covariant metric $g\lab$ of determinant $g = \text{det}(g\lab)$, and $\eta\ab=\mathrm{diag}(-1,1,1,1)$ represents an auxiliary Minkowski metric in Cartesian coordinates.} and adopting the harmonic coordinate condition $\partial_\mu h^{\alpha\mu}= 0$, we can write the ``relaxed'' Einstein field equations in the form of ordinary d'Alembert equations, namely
\begin{equation}\label{Dalembert}
\Box h\ab = \frac{16\pi G^{(d)}}{c^4} |g| \, T\ab + \Lambda\ab[h,\partial h, \partial^2 h]\,,
\end{equation}
where $\Box \equiv \eta^{\mu\nu}\partial_\mu\partial_\nu$ is the \textit{flat}-spacetime d'Alembertian operator in $D$ space-time dimensions. The gravitational source term $\Lambda\ab$ in \eqref{Dalembert} is a functional of $h^{\mu\nu}$ and its first and second space-time derivatives, and reads as
\begin{align}\label{Lambda}
    \Lambda^{\alpha\beta} =& - h^{\mu\nu} \partial_\mu\partial_\nu h^{\alpha\beta}+\partial_\mu h^{\alpha\nu} \partial_\nu h^{\beta\mu} + \frac{1}{2}g^{\alpha\beta}g_{\mu\nu}\partial_\rho h^{\mu\sigma} \partial_\sigma h^{\nu\rho} \nonumber \\ & - g^{\alpha\mu}g_{\nu\sigma}\partial_\rho h^{\beta\sigma} \partial_\mu h^{\nu\rho} -g^{\beta\mu}g_{\nu\sigma}\partial_\rho h^{\alpha\sigma} \partial_\mu h^{\nu\rho} +g_{\mu\nu}g^{\rho\sigma}\partial_\rho h^{\alpha\mu} \partial_\sigma h^{\beta\nu} \nonumber \\ & + \frac{1}{4}\bigl(2g^{\alpha\mu}g^{\beta\nu}-g^{\alpha\beta}g^{\mu\nu}\bigr)\Bigl(g_{\rho\sigma}g_{\epsilon\pi}-\frac{1}{d-1} g_{\sigma\epsilon}g_{\rho\pi}\Bigr) \partial_\mu h^{\rho\pi} \partial_\nu h^{\sigma\epsilon} \, .
\end{align}
Note the explicit dependence on the space dimension $d$ of this expression. The matter stress-energy tensor $T\ab$ will be composed of Dirac delta-functions in $d$ dimensions, say $\delta^{(d)}[\mathbf{x}-\mathbf{y}_a]$, where $\mathbf{x}$ is the field point and $\mathbf{y}_a$ are the source points labeled by $a$. Finally the $d$-dimensional gravitational constant $G^{(d)}$ is related to the usual Newton constant $G$ by
\begin{equation}\label{Geps}
G^{(d)} = G\,\ell_0^\varepsilon\,,
\end{equation}
where $\ell_0$ denotes the characteristic length associated with dimensional regularization. We shall check in Sec.~\ref{PNres} that this length scale never appears in the final 3-dimensional result.  

\subsection{Post-Newtonian metric in $d$ dimensions}
\label{metricd}

The 3PN metric is given in expanded form for general matter sources in terms of some ``elementary'' retarded potentials (sometimes called near-zone potentials) $V$, $V_i$, $K$, $\hat W_{ij}$, $\hat R_i$, $\hat X$, $\hat Z_{ij}$, $\hat Y_i$ and $\hat T$, which were introduced in Ref.~\cite{BFeom} for 3 dimensions [see Eqs.~(3.24) there] and generalized to $d$ dimensions in Ref.~\cite{BDE04}. All these potentials have a finite non-zero post-Newtonian limit when $c \rightarrow +\infty$ and parameterize the successive PN approximations. Although this decomposition in terms of near-zone potentials is convenient, such potentials have no physical meaning by themselves. Let us first define the combination
\begin{equation}\label{calV}
    \mathcal{V} \equiv V -\frac{2}{c^2} \left(\frac{d-3}{d-2}\right) K + \frac{4 \hat X}{c^4} +\frac{16 \hat T}{c^6} \, .
\end{equation}
Then the 3PN metric components can be written in the rather compact form \cite{BDE04}\footnote{This particular exponentiated form is to be consistently reexpanded at 3PN order.}
\begin{subequations}\label{metricexp}
    \begin{align}
        g^\text{PN}_{00} &= - e^{-2\mathcal{V}/c^2} \left( 1 - \frac{8 V_i V_i}{c^6} - \frac{32 \hat R_i V_i}{c^8} \right) + \mathcal{O}(c^{-10}) \, , \label{g00} \\
        g^\text{PN}_{0i} &= - e^{-\frac{(d-3)\mathcal{V}}{(d-2)c^2}} \left( \frac{4V_i}{c^3} \left[ 1 + \frac{1}{2} \left( \frac{d-1}{d-2} \frac{V}{c^2} \right)^2 \right] + \frac{8\hat R_i}{c^5} + \frac{16}{c^7} \left[ \hat Y_i + \frac{1}{2} \hat W_{ij} V_j \right] \right) + \mathcal{O}(c^{-9}) \, , \label{g0i} \\
        g^\text{PN}_{ij} &= e^{\frac{2\mathcal{V}}{(d-2)c^2}} \left( \delta_{ij} + \frac{4}{c^4} \hat W_{ij} + \frac{16}{c^6} \left[ \hat Z_{ij} - V_i V_j + \frac{1}{2(d-2)} \, \delta_{ij} V_k V_k \right] \right) + \mathcal{O}(c^{-8}) \, . \label{gij}
    \end{align}
\end{subequations}
The successive PN truncations of the field equations \eqref{Dalembert}--\eqref{Lambda} give us the equations satisfied by all the above potentials up to 3PN order. We conveniently define from the components of the matter stress-energy tensor $T\ab$ the following density, current density, and stress density
\begin{subequations}\label{sigma}
\begin{align}
    \sigma & \equiv \frac{2}{d-1} \frac{(d-2)T^{00}+T^{ii}}{c^2} \, , \\
    \sigma_i & \equiv \frac{T^{0i}}{c} \, , \\
    \sigma_{ij} & \equiv T^{ij} \, ,
\end{align}
\end{subequations}
where $T^{ii}\equiv\delta_{ij}T^{ij}$. The leading-order potentials in the metric obey
\begin{subequations}\label{potentialEq}
 \begin{align}
    \Box V = & - 4 \pi G^{(d)} \, \sigma \, , \label{dalV} \\
    \Box V_i = & - 4 \pi G^{(d)} \, \sigma_i \, , \label{dalVi} \\
    \Box K =& - 4 \pi G^{(d)} \, \sigma V\,,\label{dalK}\\
    \Box\hat W_{ij} = & - 4 \pi G^{(d)} \biggl( \sigma_{ij} - \delta_{ij} \, \frac{\sigma_{kk}}{d-2} \biggr) - \frac{1}{2} \biggl( \frac{d-1}{d-2} \biggr) \partial_i V \partial_j V \, . \label{dalWij}
    \end{align}
\end{subequations}
These potentials evidently include many PN corrections. The potentials $V$ and $V_i$ have a compact support (i.e. their source is localized on the isolated matter system) and will admit a finite limit when $\varepsilon\rightarrow 0$ without any pole. With the exception of the potential $K$ which has also a compact support,\footnote{Actually the compact-support potential $K$ does not contribute to the present calculation. Indeed, it will always be multiplied by a factor $\varepsilon=d-3$, and being compact does not generate any pole; so it does not exist in 3 dimensions.} all other potentials have, in addition to a compact-support part, a non-compact support contribution, such as that generated by the term $\propto \partial_i V \partial_j V$ in the source of $\hat W_{ij}$. This is the non-compact support piece which is the most delicate to compute because it typically generates some poles $\propto 1/\varepsilon$ at the 3PN order. The d'Alembert equations satisfied by all higher-order PN potentials, whose sources are made of non-linear combinations of lower-order potentials, are reported here for completeness:
\begin{subequations}\label{potentialEqAppA}
\begin{align}
\Box\hat R_i=&-\frac{4\pi G^{(d)}}{d-2}\left(\frac{5-d}{2}\, V \sigma_i
-\frac{d-1}{2}\, V_i\, \sigma\right)\nonumber\\
&-\frac{d-1}{d-2}\,\partial_k V\partial_i V_k -\frac{d(d-1)}{4(d-2)^2}\,\partial_t V \partial_i V\,,
\label{dalRi}\\
\Box\hat X=&-4\pi G^{(d)}\left[\frac{V\sigma_{ii}}{d-2} +2\left(\frac{d-3}{d-1}\right)\sigma_i V_i +\left(\frac{d-3}{d-2}\right)^2
\sigma\left(\frac{V^2}{2} +K\right)\right]\nonumber\\
&+\hat W_{ij}\, \partial_{ij}V +2 V_i\,\partial_t\partial_i V +\frac{1}{2}\left(\frac{d-1}{d-2}\right) V \partial^2_t V
\nonumber\\
&+\frac{d(d-1)}{4(d-2)^2}\left(\partial_t V\right)^2 -2 \partial_i V_j\, \partial_j V_i\ ,
\label{dalX}\\
\Box\hat Z_{ij}=&-\frac{4\pi G^{(d)}}{d-2}\, V\left(\sigma_{ij} -\delta_{ij}\,\frac{\sigma_{kk}}{d-2}\right) -\frac{d-1}{d-2}\, \partial_t V_{(i}\, \partial_{j)}V
+\partial_i V_k\, \partial_j V_k\nonumber\\
&+\partial_k V_i\, \partial_k V_j -2 \partial_k V_{(i}\, \partial_{j)}V_k -\frac{\delta_{ij}}{d-2}\, \partial_k V_m
\left(\partial_k V_m -\partial_m V_k\right)\nonumber\\
&-\frac{d(d-1)}{8(d-2)^3}\, \delta_{ij}\left(\partial_t V\right)^2 +\frac{(d-1)(d-3)}{2(d-2)^2}\, \partial_{(i} V\partial_{j)} K\,,
\label{dalZij}\\
\Box\hat Y_i=&-4\pi G^{(d)} \biggl[-\frac{1}{2}\left(\frac{d-1}{d-2}\right)\sigma\hat R_i -\frac{(5-d)(d-1)}{4(d-2)^2}\, \sigma V V_i +\frac{1}{2}\, \sigma_k\hat W_{ik}
+\frac{1}{2}\sigma_{ik} V_k\nonumber\\
&\hphantom{-4\pi G \biggl[}+\frac{1}{2(d-2)}\, \sigma_{kk}V_i -\frac{d-3}{(d-2)^2}\, \sigma_i \left(V^2 +\frac{5-d}{2}\, K\right)\biggr]
\nonumber\\
&+\hat W_{kl}\, \partial_{kl} V_i -\frac{1}{2}\left(\frac{d-1}{d-2}\right)
\partial_t\hat W_{ik}\, \partial_k V
+\partial_i\hat W_{kl}\, \partial_k V_l -\partial_k\hat W_{il}\, \partial_l V_k
\nonumber\\
&-\frac{d-1}{d-2}\, \partial_k V \partial_i \hat R_k -\frac{d(d-1)}{4 (d-2)^2}\, V_k\, \partial_i V \partial_k V -\frac{d(d-1)^2}{8 (d-2)^3}\, V\partial_t V\partial_i V
\nonumber\\
&-\frac{1}{2}\left(\frac{d-1}{d-2}\right)^2 V \partial_k V
\partial_k V_i
+\frac{1}{2}\left(\frac{d-1}{d-2}\right) V \partial^2_tV_i +2 V_k\, \partial_k\partial_t V_i
\nonumber\\
&+\frac{(d-1)(d-3)}{(d-2)^2}\, \partial_k K \partial_i V_k +\frac{d(d-1)(d-3)}{4(d-2)^3} \left(\partial_t V\partial_i K +\partial_i V\partial_t K\right),
\label{dalYi}\\
\Box\hat T=&-4\pi G^{(d)}\biggl[\frac{1}{2(d-1)}\, \sigma_{ij} \hat W_{ij} +\frac{5-d}{4(d-2)^2}\, V^2\sigma_{ii} +\frac{1}{d-2}\, \sigma V_i V_i
-\frac{1}{2}\left(\frac{d-3}{d-2}\right)\sigma\hat X\nonumber\\
&\hphantom{-4\pi G\biggl[}-\frac{1}{12}\left(\frac{d-3}{d-2}\right)^3 \sigma V^3 -\frac{1}{2}\left(\frac{d-3}{d-2}\right)^3 \sigma V K
+\frac{(5-d)(d-3)}{2(d-1)(d-2)}\, \sigma_i V_i V\nonumber\\
&\hphantom{-4\pi G\biggl[}+\frac{d-3}{d-1}\, \sigma_i\hat R_i -\frac{d-3}{2(d-2)^2}\, \sigma_{ii} K\biggr] +\hat Z_{ij}\, \partial_{ij}V
+\hat R_i\, \partial_t\partial_i V\nonumber\\
&-2 \partial_i V_j\, \partial_j\hat R_i -\partial_i V_j\, \partial_t\hat W_{ij} +\frac{1}{2}\left(\frac{d-1}{d-2}\right) V V_i\, \partial_t\partial_i V
+\frac{d-1}{d-2}\, V_i\, \partial_j V_i\, \partial_j V\nonumber\\
&+\frac{d(d-1)}{4(d-2)^2}\, V_i\, \partial_t V\, \partial_i V +\frac{1}{8}\left(\frac{d-1}{d-2}\right)^2 V^2\partial^2_t V
+\frac{d(d-1)^2}{8(d-2)^3}\, V\left(\partial_t V\right)^2\nonumber\\
&-\frac{1}{2}\left(\partial_t V_i\right)^2 -\frac{(d-1)(d-3)}{4(d-2)^2}\, V \partial^2_t K
-\frac{d(d-1)(d-3)}{4(d-2)^3}\, \partial_t V\, \partial_t K\nonumber\\
&-\frac{(d-1)(d-3)}{4(d-2)^2}\, K \partial^2_t V -\frac{d-3}{d-2}\, V_i\, \partial_t\partial_i K -\frac{1}{2}\left(\frac{d-3}{d-2}\right)\hat W_{ij}\,
\partial_{ij} K\,.
\label{dalT}
\end{align}
\end{subequations}

Many of the latter potentials have already been computed for compact binary systems, and we shall extensively use these results from \cite{BFeom,BDE04}. Notably, all the compact-support potentials such as $V$ and $V_i$, and all the compact-support parts of other potentials, have been computed for any field point $\mathbf{x}$, and then at the source point $\mathbf{y}_1$ following the regularization. However, the most difficult non-compact support potentials such as $\hat X$ and $\hat T$ could not be computed at any field point $\mathbf{x}$, and were regularized directly on the particle's world-line. Since for the equations of motion we needed only the \textit{gradients} of these potentials, only the gradients were regularized on the particle, yielding the results for $(\partial_i\hat X)(\mathbf{y}_1)$ and $(\partial_i\hat T)(\mathbf{y}_1)$ needed in the equations of motion. However the 3PN metric requires the values of the potentials themselves regularized on the particles, i.e. $\hat X(\mathbf{y}_1)$ and $\hat T(\mathbf{y}_1)$. For the present work we have therefore to compute, using the tools developed in \cite{BFeom,BDE04}, the difficult non-linear potentials $\hat X(\mathbf{y}_1)$ and $\hat T(\mathbf{y}_1)$, and especially the non-compact support parts therein. Unfortunately, the potential $\hat X$ is always the most tricky to compute, because its source involves the cubically-non-linear and non-compact-support term $\hat W_{ij}\, \partial_{ij}V$, and it has to be evaluated at relative 1PN order.

In this calculation we also meet a new difficulty with respect to the computation of the 3PN equations of motion. Indeed, we find that the potential $\hat X$ is divergent because of the bound of the Poisson-like integral at \textit{infinity}.\footnote{However the other potential $\hat T$, which is merely Newtonian, is convergent at infinity.} Thus the potential $\hat X$ develops an IR divergence, in addition to the UV divergence due to the singular nature of the source and which is cured by dimensional regularization. The IR divergence is a particular case of the well-known divergence of Poisson integrals in the PN expansion for general (regular) sources, linked to the fact that the PN expansion is a singular perturbation expansion, with coefficients typically blowing up at spatial infinity. The IR divergence will be treated in Sec.~\ref{IR}.

The 3PN metric \ref{metricexp} is valid for a general isolated matter system, and we apply it to the case of a system of $N$ point-particles with ``Schwarzschild'' masses $m_a$ and without spins (here $a=1,\cdots,N$). In this case we have
\begin{subequations}
\begin{align}
\sigma (\mathbf{x} , t) &= \sum_a \tilde\mu_a  \, \delta^{(d)} [\mathbf{x} - \mathbf{y}_a (t)]\,,\\\sigma_{i} (\mathbf{x} , t) &= \sum_a \mu_a \,v_a^i\, \delta^{(d)} [\mathbf{x} - \mathbf{y}_a (t)]\,,\\\sigma_{ij} (\mathbf{x} , t) &= \sum_a \mu_a \,v_a^i\,v_a^j \,\delta^{(d)} [\mathbf{x} - \mathbf{y}_a (t)]\,,
\end{align}
\end{subequations}
where $\delta^{(d)}$ denotes the Dirac density in $d$ spatial dimensions such that $\int \ud^d \mathbf{x} \, \delta^{(d)}(\mathbf{x}) = 1$. We defined the effective time-varying masses of the particles by
\begin{equation}
\mu_a (t) = \frac{m_a}{\sqrt{(gg_{\alpha \beta}) (\mathbf{y}_a, t) \, v_a^{\alpha} \, v_a^{\beta}/c^2}} \,,
\end{equation}
together with $\tilde\mu_a = \frac{2}{d-1} \left[ d - 2 + \mathbf{v}_a^2/c^2 \right] \mu_a$.

\subsection{Dimensional regularization of Poisson integrals}
\label{reg}

In the PN approximation we break the hyperbolic d'Alembertian operator $\Box$ in Eqs.~\eqref{potentialEq}--\eqref{potentialEqAppA} into the elliptic Laplacian operator $\Delta$ and the small PN retardation term $c^{-2}\partial_t^2$, which is put in the RHS of the equation and iterated. Neglecting the radiation-reaction effects, this means that we solve the d'Alembert equations by means of the symmetric Green function
\begin{equation}\label{sym}
\Box^{-1}_\mathrm{sym} = \Delta^{-1}+\frac{1}{c^2}\Delta^{-2}\partial_t^2 + \mathcal{O}(c^{-4}) \, .
\end{equation}
We consider only the 1PN retardation because the potentials $\hat X$ and $\hat T$ which are the only ones to be computed are to be evaluated at 1PN order at most. We are thus led to define the dimensional regularization of Poisson or twice-iterated Poisson integrals. 

Let $F(\mathbf{x})$ be the generic form of the functions representing the PN potentials in $d$ dimensions. For simplicity we shall treat only the case of the non-compact support terms. Compact support potentials or compact part of potentials, such as $V$ or the first term in the source of $\hat W_{ij}$ do not generate poles in $d$ dimensions and were dealt with in Ref.~\cite{BDE04}. Also, we consider only Poisson integrals and refer to \cite{BDE04} for the procedure for iterated Poisson integrals. So we want to compute a typical Poisson potential
\begin{equation}\label{Px'}
P(\mathbf{x}') = \Delta^{-1}\left[F(\mathbf{x})\right] \equiv -\frac{k}{4\pi}\int \ud^d{\mathbf{x}}\,\frac{F({\mathbf{x}})}{\vert{\mathbf{x}}-\mathbf{x}'\vert^{d-2}}\,.
\end{equation}
We employ the Green function $u=k\,\vert\mathbf{x}\vert^{2-d}$ of the Laplace operator in $d$-dimensions, satisfying $\Delta u = - 4\pi\,\delta^{(d)}(\mathbf{x})$, where
\begin{equation}\label{k}
k \equiv \frac{\Gamma\left(\frac{d-2}{2}\right)}{\pi^{\frac{d-2}{2}}}
\end{equation}
is defined from the standard Eulerian gamma function.\footnote{The constant $k$ tends to $1$ when $d\rightarrow 3$, and was formerly denoted $\tilde{k}$ in Ref.~\cite{BDE04}.} Furthermore we want to evaluate the Poisson integral $P(\mathbf{x}')$ on one of the singular points, say $\mathbf{x}'=\mathbf{y}_a$. As we shall see the Poisson potential we have to deal with will not only be divergent on the singularities but also at infinity, i.e. when the source point $r \equiv \vert \mathbf{x} \vert \rightarrow +\infty$. To delineate these problems we introduce a constant radius $\mathcal{R}>0$, and split the Poisson potential into a near-zone integral corresponding to  $r<\mathcal{R}$, and a far-zone one such that $r>\mathcal{R}$:
\begin{equation}\label{NZFZ}
    P(\mathbf{x}') = P_<(\mathbf{x}') +  P_>(\mathbf{x}')\,.
\end{equation}
The near-zone integral $P_<$ will contain the local or ultra-violet (UV) singularities, due to the singular nature of the point-mass source, while the far-zone integral $P_>$ will have an infra-red (IR) divergence at infinity, which is actually a general feature of the PN expansion for any post-Newtonian source.

\subsubsection{UV divergence}
\label{UV}

The UV divergences will be dealt with using dimensional regularization. Non-compact support terms are generated by a generic function $F(\mathbf{x})$ which extends to all space. For all needed cases we can assume that $F$ is smooth everywhere except at the singular points $\mathbf{y}_a$, around which it admits a singular expansion in powers of $r_a\equiv\vert\mathbf{x}-\mathbf{y}_a\vert$ when $r_a\rightarrow 0$, of the type
\begin{equation}\label{Fdx}
F(\mathbf{x})=\sum_{p=p_0}^P \sum_{q=q_0}^{q_1}r_a^{p+q\varepsilon}\mathop{f}_a{}_{\!\!p,q}(\mathbf{n}_a) + o\bigl(r_a^P\bigr)\,,
\end{equation}
for any $P\in\N$. The coefficients ${}_af_{p,q}(\mathbf{n}_a)$ are functions of the unit direction $\mathbf{n}_a = (\mathbf{x}-\mathbf{y}_a)/r_a$, and depend on the dimension through $\varepsilon = d-3$, and also on the DR length scale $\ell_0$.\footnote{More precisely, the coefficients depend on $\ell_0$ as ${}_af_{p,q} \propto \ell_0^{-q \varepsilon}$, as can be seen from the expansion \eqref{Fdx}.} The powers of $r_a$ are of the type $p+q\varepsilon$, where $p$ and $q$ are relative integers ($p,\,q\in\Z$) with values limited as indicated. The singular expansion \eqref{Fdx} will yield some UV-type divergence of the Poisson potential \eqref{Px'}. Relying on analytic continuation, we can evaluate $F$ at the location of particle $a$ simply by taking the limit $\mathbf{x} \rightarrow \mathbf{y}_a$. Indeed, we can check that the dimension $d$ can always be chosen such that $F$ is non-singular in this limit. Thus,
\begin{equation}\label{Fya}
    F(\mathbf{y}_a) = \mathrm{AC}\Bigl[\lim_{\mathbf{x} \rightarrow \mathbf{y}_a} F(\mathbf{x})\Bigr] \,,
\end{equation}
and we may then consider the expansion when $\varepsilon\rightarrow 0$. From now on the analytic continuation process will be implicitly assumed without indication.

The near-zone part of the Poisson integral of the generic function $F$ outside the singularities is defined by
\begin{equation}\label{Pdx}
P_<(\mathbf{x}')= -\frac{k}{4\pi}\int_{r<\mathcal{R}}\ud^d{\mathbf{x}}\,\frac{F({\mathbf{x}})}{\vert{\mathbf{x}}-\mathbf{x}'\vert^{d-2}}\,,
\end{equation}
in which the upper bound of the integral is set at the intermediate radius $\mathcal{R}$. The singular behavior of this integral at the source points, i.e. when $\mathbf{x} = \mathbf{y}_a$, is automatically taken care of by dimensional continuation down to $d=3$. Next we evaluate the integral at the singular point $\mathbf{x}' = \mathbf{y}_a$ itself. The result is easy in DR,\footnote{This is in contrast with the difficult formulation necessary in Hadamard's regularization; see Refs.~\cite{BFeom,BDE04}.} as we are allowed to simply replace $\mathbf{x}'$ by $\mathbf{y}_a$ into \eqref{Pdx}. Thus,
\begin{equation}\label{Pda}
P_<(\mathbf{y}_a)= -\frac{k}{4\pi}\int_{r<\mathcal{R}}\ud^d{\mathbf{x}}\,\frac{F(\mathbf{x})}{r_a^{d-2}}\,,
\end{equation}
which is the main result of DR, as applied to UV divergences.

In practical calculations we are interested in the three-dimensional limit, so we perform the Laurent expansion of the previous result when $\varepsilon =
d-3 \rightarrow 0$. As we know from previous work \cite{BDE04}, the expression \eqref{Pda} is finite for any of the non-compact potentials up to 2.5PN order, but will develop a simple pole $\propto 1/\varepsilon$ at the 3PN order. The poles correspond to the occurrence of logarithmic divergences in the three-dimensional calculation \cite{BFeom}, and are in fact associated with our particular coordinate choice. Indeed, similar calculations performed at the 3PN level in ADM-like coordinates within DR are pole-free \cite{DJSdim}. The information we shall need is the pole part followed by the associated finite part when $\varepsilon\rightarrow 0$; we shall usually leave aside the remainder term $\mathcal{O}(\varepsilon)$.\footnote{Although those remainders $\mathcal{O}(\varepsilon)$ present in Newtonian terms will be kept because they might get multiplied by some poles $\varepsilon^{-1}$ at the 3PN order, therefore yielding finite contributions at 3PN order.} We thus consider the expansion
\begin{equation}\label{PdAexp}
P_<(\mathbf{y}_a) = \frac{1}{\varepsilon}P_<^{(-1)}(\mathbf{y}_a)+P_<^{(0)}(\mathbf{y}_a)+
 \mathcal{O}(\varepsilon)\,,
\end{equation}
and we look for the pole-part $P_<^{(-1)}(\mathbf{y}_a)$ and finite-part $P_<^{(0)}(\mathbf{y}_a)$ coefficients. Beware that our terminology is slightly misleading, because we shall conveniently include in the pole part $P_<^{(-1)}(\mathbf{y}_a)$ some dependence in $\varepsilon$, which will of course be thought as being expanded when $\varepsilon\rightarrow 0$ up to first order in $\varepsilon$, therefore yielding a finite contribution $\mathcal{O}(\varepsilon^0)$ to be added to the finite part $P_<^{(0)}(\mathbf{y}_a)$. Combining previous results in Sec.~IV of \cite{BDE04}, we find that the pole part is explicitly given by
\begin{equation}\label{polepart}
P_<^{(-1)}(\mathbf{y}_a)= -\frac{1}{1+\varepsilon}\sum_{q=q_0}^{q_1}\Biggl[\frac{1}{q}\langle\mathop {f}_a{}_{\!\!-2,q}\rangle + \frac{1}{q+1}\sum_{b\not=a}\sum_{\ell=0}^{+\infty}\frac{(-)^\ell}{\ell!}\partial_{i_1 \cdots i_\ell} \biggl(\frac{1}{r_{ab}^{1+\varepsilon}}\biggr)\langle n_b^{i_1 \cdots i_\ell}\,\mathop{f}_b{}_{\!\!-(\ell+3),q}\rangle\Biggr] \,.
\end{equation}
The first term is the contribution of the singularity $a$ which is clearly singled out, while the second term comes from all the other singularities $b \neq a$. The bracket notation in \eqref{polepart} refers to the angular average performed in $d$ dimensions, i.e.
\begin{equation}\label{anguld}
\langle\mathop{f}_a{}_{\!\!p,q}\rangle \equiv\int\frac{\ud\Omega_{d-1}(\mathbf{n}_a)}{\Omega_{d-1}} \mathop{f}_a{}_{\!\!p,q}(\mathbf{n}_a)\,,
\end{equation}
where the volume $\Omega_{d-1}$ of the $(d-1)$-dimensional sphere is given by
\begin{equation}\label{Omegad}
\Omega_{d-1} = \frac{2\pi^{\frac{d}{2}}}{\Gamma\left(\frac{d}{2}\right)} = \frac{4\pi}{k\,(d-2)}\,.
\end{equation}
We observe that the pole part \eqref{polepart} depends on the behavior of the function $F$ only for $\mathbf{x}$ in a neighborhood of the singularities,
through the singular expansion coefficients ${}_af_{p,q}$ with $p \leqslant -2$. The result for the pole part heavily relies on DR, and depends on the DR scale $\ell_0$ introduced in \eqref{Geps}. However, as it is ``localized'' on the singularities, the pole part is independent of the radius $\mathcal{R}$.

On the contrary, the finite part coefficient $P_<^{(0)}(\mathbf{y}_a)$ depends on all the ``bulk'' of the integration outside the particle's world-lines and in particular on the radius $\mathcal{R}$. This finite part essentially corresponds to what we would naively compute in 3 dimensions, i.e. by simply replacing $d=3$ into \eqref{Pda}; the result, however, would be ill-defined as it stands. In DR the finite part coefficient in \eqref{PdAexp} has a precise meaning, and we find that it agrees with the so-called Hadamard \textit{partie finie} integral \cite{Hadamard,Schwartz}
\begin{equation}\label{partiefinie}
P_<^{(0)}(\mathbf{y}_a)= -\frac{1}{4\pi} \mathrm{Pf}_{\ell_0} \int_{r<\mathcal{R}}\ud^3{\mathbf{x}}\,\frac{F^{(0)}({\mathbf{x}})}{r_a} \,,
\end{equation}
where $F^{(0)}$ is the function $F$ computed with $\varepsilon=0$. Here ``$ \mathrm{Pf}_{\ell_0}$'' stands for the partie finie which depends on the arbitrary scale $\ell_0$ playing here the role of the Hadamard regularization scales. Thus, all the Hadamard regularization scales, one for each particles (they were previously denoted $s_1,\cdots,s_N$ in \cite{BFeom}), are to be replaced by the unique scale $\ell_0$. For instance, in the equivalent representation of Hadamard's partie finie as an analytic continuation, making explicit the presence of those arbitrary constant scales, we have
\begin{equation}\label{partiefinie2}
P_<^{(0)}(\mathbf{y}_a)= -\frac{1}{4\pi}\mathrm{FP}\int_{r<\mathcal{R}}\ud^3{\mathbf{x}}
  \,\left(\frac{r_1}{\ell_0}\right)^{\!\!\alpha_1}\!\!\cdots\left(\frac{r_N}{\ell_0}\right)^{\!\!\alpha_N}
  \frac{F^{(0)}({\mathbf{x}})}{r_a} \,,
\end{equation}
where the symbol FP is understood as the finite part of the Laurent expansion of the integral when all of the $\alpha_a$'s tend to zero. The Hadamard partie finie \eqref{partiefinie} or \eqref{partiefinie2} is extremely convenient to implement in practical computations.

\subsubsection{IR divergence}
\label{IR}

Next we have also to worry about the IR-type divergence of the Poisson potential $P$, due to the behavior of the source $F$ at spatial infinity, when $r \rightarrow +\infty$. Indeed, we find that the near-zone potential $\hat X$ we have to evaluate (and which is to be computed at 1PN relative order) is given by an iterated Poisson integral which is divergent at infinity. The appearance of a divergent near-zone potential $\hat X(\mathbf{y}_a)$ is a novel feature of the present calculation; indeed the problem did not arise in the previous computation of the 3PN equations of motion because we needed instead the \textit{gradient} $(\partial_i\hat X)(\mathbf{y}_a)$, which is convergent. 

Fortunately, the problem of IR divergences has been solved in the general case, for any isolated PN source and at any PN order. Here we shall follow the formalism of Ref.~\cite{PB02} which uses systematically a regularized version of the Poisson integral which is appropriate for solving the hierarchy of PN equations. The idea is to introduce inside the Poisson integral a regularization factor $(r/r_0)^B$, where $B$ is a complex number, and where $r_0$ is an arbitrary IR scale (as $\ell_0$ is an arbitrary UV scale). The regularized Poisson integral is then defined as the finite part (FP) coefficient, i.e. the coefficient of the zeroth power of $B$, in the Laurent expansion of the integral when $B\rightarrow 0$. It was proved in~\cite{PB02} that the latter regularized Poisson integral is a solution of the Poisson equation for a general regular (smooth) source, and is amenable to iteration up to any PN order.

In principle the latter procedure is defined in $3$ dimensions. However since we are here solving the $d$-dimensional field equations, we shall first define it in $d$ dimensions, so that the far-zone part of the Poisson potential reads
\begin{equation}\label{PFd}
P_>(\mathbf{x}')= -\frac{k}{4\pi}\mathrm{FP}\int_{r>\mathcal{R}}\ud^d{\mathbf{x}}\,\left(\frac{r}{r_0}\right)^{\!\!B}\frac{F(\mathbf{x})}{\vert{\mathbf{x}}-\mathbf{x}'\vert^{d-2}} \,,
\end{equation}
where FP refers to the finite part when $B\rightarrow 0$. The precise meaning of considering the FP process on a $d$-dimensional integral has been discussed in \cite{BDEI05dr}.\footnote{It could be possible to use dimensional regularization to cure not only the UV divergences but also the IR ones (without the FP when $B\rightarrow 0$). However this would imply major modifications of the PN iteration scheme for a general source; this has not been attempted.} Here, since we consider only the far-zone part of the integral free of UV divergences, we can immediately take the limiting case $\varepsilon=0$ and get
\begin{equation}\label{PFd3}
P_>^{(0)}(\mathbf{x}')= -\frac{1}{4\pi}\mathop{\mathrm{FP}}\int_{r>\mathcal{R}}\ud^3{\mathbf{x}}\,\left(\frac{r}{r_0}\right)^{\!\!B}\frac{F^{(0)}(\mathbf{x})}{\vert{\mathbf{x}}-\mathbf{x}'\vert} \,.
\end{equation}
The result will depend both on the IR cut-off scale $r_0$ and intermediate radius $\mathcal{R}$, but we shall check that these constants disappear in the final results. At the point $\mathbf{y}_a$ we have
\begin{equation}\label{PFd3a}
P_>^{(0)}(\mathbf{y}_a)= -\frac{1}{4\pi}\mathrm{FP}\int_{r>\mathcal{R}}\ud^3{\mathbf{x}}\,\left(\frac{r}{r_0}\right)^{\!\!B}\frac{F^{(0)}(\mathbf{x})}{r_a} \,.
\end{equation}

Finally, the sum of the near-zone integral \eqref{partiefinie2} and far-zone one \eqref{PFd3a} gives our complete prescription for the finite part of the dimensionally regularized Poisson integral as
\begin{equation}\label{PFd3atotal}
P^{(0)}(\mathbf{y}_a)= -\frac{1}{4\pi}\mathrm{FP} \int \ud^3{\mathbf{x}} \, \left(\frac{r}{r_0}\right)^{\!\!B} \left(\frac{r_1}{\ell_0}\right)^{\!\!\alpha_1}\!\!\cdots\left(\frac{r_N}{\ell_0}\right)^{\!\!\alpha_N} \frac{F^{(0)}(\mathbf{x})}{r_a} \, .
\end{equation}
We do not detail how this integral is computed in practice but refer to previous works \cite{BFeom} and, for the treatment of the bound at infinity, Sec.~IV~C of \cite{BI04mult}. We checked that the sum of the resulting UV and IR-regularized Poisson integral is independent of the arbitrary constant length-scale $\mathcal{R}$.

\section{Post-Newtonian results}
\label{PNres}

\subsection{The regularized 3PN metric}
\label{PNmetric}

The post-Newtonian metric is generated by a system of two point particles, and computed at the location of the particle 1 following the prescription \eqref{gPNdr}. Here we shall somewhat abusively simply denote this metric $g\lab(\mathbf{y}_1,t)$, or in short $g\lab(y_1)$, and similarly for other quantities evaluated at the location of particle 1, so that
\begin{align}\label{gab1}
    g\lab(y_1) \equiv \mathop{\mathrm{AC}}_{\varepsilon\rightarrow 0}\,\Bigl[\lim_{\mathbf{x}\rightarrow\mathbf{y}_1} g^\text{PN}\lab(\mathbf{x},t)\Bigr]\,.
\end{align}
We compute all the required near-zone potentials $V$, $V_i$, $\cdots$, $\hat X$, $\hat T$ at point 1 (actually, only $\hat X$ and $\hat T$ are new in the present computation with respect to Ref.~\cite{BDE04}) and regularize them according to the procedure of the previous section.
The regularized metric in harmonic coordinates is now obtained in closed analytic form up to 3PN order as\footnote{The masses $m_1$ and $m_2$ have an arbitrary mass ratio. As usual we denote by $r_{12} = \vert \mathbf{y}_1 - \mathbf{y}_2 \vert$ the relative distance between the two particles in harmonic coordinates, by $\mathbf{n}_{12} = (\mathbf{y}_1 - \mathbf{y}_2)/ r_{12}$ the direction from particle $2$ to particle $1$, and by $\mathbf{v}_{12} = \mathbf{v}_1 - \mathbf{v}_2$ the relative velocity, where $\mathbf{v}_a = \ud \mathbf{y}_a / \ud t$ is the coordinate velocity of particle $a$. The Euclidean scalar product between two vectors $\mathbf{A}$ and $\mathbf{B}$ is $(A B)$. Parentheses around indices are used to indicate symmetrization, i.e. $A^{(i} B^{j)} \equiv \frac{1}{2} ( A^i B^j + A^j B^i)$.}
\begin{subequations}\label{g1dr}
    \begin{align}
        g_{00}(y_1) &= -1 + \frac{2 G m_2}{c^2 r_{12}} + \frac{G m_2}{c^4 r_{12}} \biggl[ 4 v_2^2 - (n_{12}v_2)^2 - 3 \frac{G m_1}{r_{12}} - 2 \frac{G m_2}{r_{12}} \biggr] \nonumber \\ &+ \frac{G m_2}{c^6 r_{12}} \biggl[\frac{3}{4} (n_{12}v_2)^4 - 3 (n_{12}v_2)^2 v_2^2 + 4 v_2^4 + \frac{G m_2}{r_{12}} \biggl( 3 (n_{12}v_2)^2 - v_2^2 + 2 \frac{G m_2}{r_{12}} \biggr) \nonumber \\ &+ \frac{G m_1}{r_{12}} \biggl( - \frac{87}{4} (n_{12}v_1)^2 + \frac{47}{2} (n_{12}v_1)(n_{12}v_2) - \frac{55}{4} (n_{12}v_2)^2 - \frac{39}{2} (v_1 v_2) + \frac{23}{4} v_1^2 \nonumber \\ &+\frac{47}{4} v_2^2 - \frac{G m_1}{r_{12}} + \frac{17}{2} \frac{G m_2}{r_{12}} \biggr) \biggr] + \frac{G m_2}{c^8 r_{12}} \biggl[ - \frac{5}{8} (n_{12}v_2)^6 - 5 (n_{12}v_2)^2 v_2^4 + 3 (n_{12}v_2)^4 v_2^2 \nonumber \\ &+ 4 v_2^6 + \frac{G m_2}{r_{12}} \biggl( -4 (n_{12}v_2)^4 + 5 (n_{12}v_2)^2 v_2^2 - v_2^4 \biggr) + \frac{G m_1}{r_{12}} \biggl( - \frac{617}{24} (n_{12}v_1)^4 \nonumber \\ &+ \frac{491}{6} (n_{12}v_1)^3 (n_{12}v_2) - \frac{225}{4} (n_{12}v_1)^2 (n_{12}v_2)^2  + \frac{41}{2} (n_{12}v_1) (n_{12}v_2)^3 + \frac{53}{8} (n_{12}v_2)^4 \nonumber \\ &- \frac{79}{4} (n_{12}v_1)^2 (v_1 v_2) + 42 (n_{12}v_1) (n_{12}v_2) (v_1 v_2) + \frac{101}{4} (n_{12}v_2)^2 (v_1 v_2) + \frac{49}{4} (v_1 v_2)^2 \nonumber \\ &- \frac{27}{8} (n_{12}v_1)^2 v_1^2 + \frac{23}{4} (n_{12}v_1) (n_{12}v_2) v_1^2 - \frac{273}{8} (n_{12}v_2)^2 v_1^2 - 25 (v_1 v_2) v_1^2 + \frac{39}{8} v_1^4 \nonumber \\ &- \frac{305}{8} (n_{12}v_1)^2 v_2^2 + \frac{139}{4} (n_{12}v_1) (n_{12}v_2) v_2^2 - \frac{291}{8} (n_{12}v_2)^2 v_2^2 - 62 (v_1 v_2) v_2^2 + \frac{77}{2} v_1^2 v_2^2 \nonumber \\ &+ \frac{235}{8} v_2^4 \biggr) + \frac{G^2 m_1^2}{r_{12}^2} \biggl( \ln{\left( \frac{r_{12}}{r_0} \right)} \biggl\{ 32 (n_{12}v_{12})^2 - \frac{32}{3} v_{12}^2 \biggr\} + (n_{12}v_1)^2 \biggl\{ \frac{12021}{100} + \frac{182}{5 \varepsilon} \nonumber \\ &- \frac{546}{5} \ln{\left( \frac{r_{12} \, p}{\ell_0} \right)} \biggr\} + (n_{12}v_1) (n_{12}v_2) \biggl\{ - \frac{2873}{50} - \frac{292}{5 \varepsilon} + \frac{876}{5} \ln{\left( \frac{r_{12} \, p}{\ell_0} \right)} \biggr\} \nonumber \\ &+ (n_{12}v_2)^2 \biggl\{ - \frac{21}{4} + \frac{22}{\varepsilon} - 66 \ln{\left( \frac{r_{12} \, p}{\ell_0} \right)} \biggr\} + (v_1 v_2) \biggl\{ \frac{16349}{450} + \frac{292}{15 \varepsilon} \nonumber \\ &- \frac{292}{5} \ln{\left( \frac{r_{12} \, p}{\ell_0} \right)} \biggr\} + v_1^2 \biggl\{ - \frac{38573}{900} - \frac{182}{15 \varepsilon} + \frac{182}{5} \ln{\left( \frac{r_{12} \, p}{\ell_0} \right)} \biggr\} + v_2^2 \biggl\{ - \frac{125}{36} \nonumber \\ &- \frac{22}{3 \varepsilon} + 22 \ln{\left( \frac{r_{12} \, p}{\ell_0} \right)} \biggr\} + \frac{G m_1}{r_{12}} \biggl\{ \frac{719}{36} + \frac{22}{3 \varepsilon} - \frac{88}{3} \ln{\left( \frac{r_{12} \, p}{\ell_0} \right)} + \frac{32}{3} \ln{\left( \frac{r_{12}}{r_0} \right)} \biggr\} \biggr) \nonumber \\ &+ \frac{G^2 m_1 m_2}{r_{12}^2} \biggl( \ln{\left( \frac{r_{12}}{r_0} \right)} \biggl\{ 32 (n_{12}v_{12})^2 - \frac{32}{3} v_{12}^2 \biggr\} + (n_{12}v_1)^2 \biggl\{ - \frac{109}{12} - \frac{141}{16} \pi^2 \biggr\} \nonumber \\ &+ (n_{12}v_1) (n_{12}v_2) \biggl\{ - \frac{197}{6} + \frac{177}{8} \pi^2 \biggr\} + (n_{12}v_2)^2 \biggl\{ \frac{391}{6} - \frac{213}{16} \pi^2 \biggr\} + (v_1 v_2) \biggl\{ \frac{812}{9} \nonumber \\ &- \frac{59}{8} \pi^2 \biggr\} + v_1^2 \biggl\{ - \frac{299}{18} + \frac{47}{16}\pi^2 \biggr\} + v_2^2 \biggl\{ - \frac{1097}{18} + \frac{71}{16}\pi^2 \biggr\} + \frac{G m_1}{r_{12}} \biggl\{ - \frac{769}{36} - \frac{15}{8} \pi^2 \nonumber \\ &+ \frac{22}{3 \varepsilon} - \frac{88}{3} \ln{\left( \frac{r_{12} \, p}{\ell_0} \right)} + \frac{64}{3} \ln{\left( \frac{r_{12}}{r_0} \right)} \biggr\} + \frac{G m_2}{r_{12}} \biggl\{ - \frac{586}{9} - \frac{15}{8} \pi^2 - \frac{22}{3 \varepsilon} \nonumber \\ &+ \frac{88}{3} \ln{\left( \frac{r_{12} \, p}{\ell_0} \right)} + \frac{32}{3} \ln{\left( \frac{r_{12}}{r_0} \right)} \biggr\} \biggr) + \frac{G^2 m_2^2}{r_{12}^2} \biggl( - (n_{12}v_2)^2 + 2 v_2^2 - 2 \frac{G m_2}{r_{12}} \biggr) \biggr] \nonumber \\ &+ \mathcal{O}(c^{-10}) \, , \label{g001dr} \\
        g_{0i}(y_1) &= - \frac{4 G m_2}{c^3 r_{12}} v_2^i + \frac{G m_2}{c^5 r_{12}} \biggl[ v_2^i \biggl( 2 (n_{12}v_2)^2 - 4 v_2^2 - 2 \frac{G m_1}{r_{12}} + \frac{G m_2}{r_{12}} \biggr) + 4 \frac{G m_1}{r_{12}} v_1^i \nonumber \\ &+ n_{12}^i \biggl( \frac{G m_1}{r_{12}} \bigl\{ 10 (n_{12}v_1) + 2 (n_{12}v_2) \bigr\} - \frac{G m_2}{r_{12}} (n_{12}v_2) \biggr) \biggr] + \frac{G m_2}{c^7 r_{12}} \biggl[ v_2^i \biggl( - \frac{3}{2} (n_{12}v_2)^4 \nonumber \\ &+ 4 (n_{12}v_2)^2 v_2^2 - 4 v_2^4 + \frac{G m_1}{r_{12}} \biggl\{ 48 (n_{12}v_1)^2 - 44 (n_{12}v_1) (n_{12}v_2) + 10 (n_{12}v_2)^2 \nonumber \\ &+ 40 (v_1 v_2) - 16 v_2^2 - 26 v_1^2 \biggr\} + \frac{G m_2}{r_{12}} \biggl\{ - 2 (n_{12}v_2)^2 + v_2^2 - 2 \frac{G m_2}{r_{12}} \biggr\} \nonumber \\ &+ \frac{G^2 m_1 m_2}{r_{12}^2} \biggl\{ \frac{95}{6} - \frac{3}{4} \pi^2 \biggr\} + \frac{G^2 m_1^2}{r_{12}^2} \biggl\{ - \frac{1102}{75} - \frac{12}{5 \varepsilon} + \frac{36}{5} \ln{\left( \frac{r_{12} \, p}{\ell_0} \right)} \biggr\} \biggr) \nonumber \\ &+ v_1^i \biggl( \frac{G m_1}{r_{12}} \biggl\{ - \frac{17}{2} (n_{12}v_1)^2 - 15 (n_{12}v_1) (n_{12}v_2) + \frac{43}{2} (n_{12}v_2)^2 + 3 (v_1 v_2) + \frac{17}{2} v_1^2 \nonumber \\ &- \frac{15}{2} v_2^2 \biggr\} + \frac{G^2 m_1 m_2}{r_{12}^2} \biggl\{ - \frac{57}{2} + \frac{3}{4} \pi^2 \biggr\} + \frac{G^2 m_1^2}{r_{12}^2} \biggl\{ \frac{1852}{75} + \frac{12}{5 \varepsilon} - \frac{36}{5} \ln{\left( \frac{r_{12} \, p}{\ell_0} \right)} \biggr\} \biggr) \nonumber \\ &+ n_{12}^i \biggl( \frac{G m_1}{r_{12}} \biggl\{ \frac{21}{2} (n_{12}v_1)^3 - \frac{43}{2} (n_{12}v_1)^2 (n_{12}v_2) - \frac{29}{2} (n_{12}v_1) (n_{12}v_2)^2 + \frac{3}{2} (n_{12}v_2)^3 \nonumber \\ &- 11 (n_{12}v_1) (v_1 v_2) - 19 (n_{12}v_2) (v_1 v_2) + \frac{1}{2} (n_{12}v_1) v_1^2 + \frac{39}{2} (n_{12}v_2) v_1^2 + \frac{41}{2} (n_{12}v_1) v_2^2 \nonumber \\ &+ \frac{3}{2} (n_{12}v_2) v_2^2 \biggr\} + \frac{G m_2}{r_{12}} (n_{12}v_2) \biggl\{ 2 (n_{12}v_2)^2 - v_2^2 - 2 \frac{G m_2}{r_{12}} \biggr\} + \frac{G^2 m_1 m_2}{r_{12}^2} \biggl\{ \frac{51}{2} (n_{12}v_1) \nonumber \\ &- \frac{97}{2} (n_{12}v_2) - \frac{9}{4} \pi^2 (n_{12}v_{12}) \biggr\} + \frac{G^2 m_1^2}{r_{12}^2} \biggl\{ - \frac{2337}{25} (n_{12}v_1) + \frac{1237}{25} (n_{12}v_2) - \frac{36}{5 \varepsilon} (n_{12}v_{12}) \nonumber \\ &+ \frac{108}{5} (n_{12}v_{12}) \ln{\left( \frac{r_{12} \, p}{\ell_0} \right)} \biggr\} \biggr) \biggr] + \mathcal{O}(c^{-9}) \, , \label{g0i1dr} \\
        g_{ij}(y_1) &= \delta^{ij} + \frac{2 G m_2}{c^2 r_{12}} \delta^{ij} + \frac{G m_2}{c^4 r_{12}} \biggl[ \delta^{ij} \biggl( - (n_{12}v_2)^2 + \frac{G m_1}{r_{12}} + \frac{G m_2}{r_{12}} \biggr) + 4 v_2^i v_2^j \nonumber \\ &+ n_{12}^i n_{12}^j \biggl( - 8 \frac{G m_1}{r_{12}} + \frac{G m_2}{r_{12}} \biggr) \biggr] + \frac{G m_2}{c^6 r_{12}} \biggl[ \delta^{ij} \biggl( \frac{3}{4} (n_{12}v_2)^4 - v_2^2 (n_{12}v_2)^2 \nonumber \\ &- \frac{G m_2}{r_{12}} (n_{12}v_2)^2 + \frac{G m_1}{r_{12}} \biggl\{ - \frac{71}{4} (n_{12}v_{12})^2 + \frac{47}{4} v_{12}^2 - 9 \frac{G m_1}{r_{12}} + \frac{25}{6} \frac{G m_2}{r_{12}} \biggr\} \biggr) \nonumber \\ &+ v_2^i v_2^j \biggl( 4 v_2^2 - 2 (n_{12}v_2)^2 - 10 \frac{G m_1}{r_{12}} - \frac{G m_2}{r_{12}} \biggr) + 24 \frac{G m_1}{r_{12}} v_1^{(i} v_2^{j)} - 16 \frac{G m_1}{r_{12}} v_1^i v_1^j \nonumber \\ &+ n_{12}^i n_{12}^j \biggl( \frac{G m_1}{r_{12}} \biggl\{ -16 (n_{12}v_1)^2 + 32 (n_{12}v_1) (n_{12}v_2) - 12 v_{12}^2 + 28 \frac{G m_1}{r_{12}} \biggr\} \nonumber \\ &+ \frac{G m_2}{r_{12}} \biggl\{ -2 (n_{12}v_2)^2 + 3 \frac{G m_1}{r_{12}} + 2 \frac{G m_2}{r_{12}} \biggr\} \biggr) + 40 \frac{G m_1}{r_{12}} (n_{12}v_{12}) \, n_{12}^{(i} v_1^{j)} \nonumber \\ &+ n_{12}^{(i} v_2^{j)} \biggl( \frac{G m_1}{r_{12}} \bigl\{ - 60 (n_{12}v_1) + 36 (n_{12}v_2) \bigr\} + 2 \frac{G m_2}{r_{12}} (n_{12}v_2) \biggr) \biggr] + \mathcal{O}(c^{-8}) \, . \label{gij1dr}
    \end{align}
\end{subequations}
We indicate explicitly the post-Newtonian remainders $\mathcal{O}(c^{-n})$. This metric agrees up to 2PN order with the already known result obtained in \cite{BFP98}, and recently used in \cite{Det08} for obtaining the self-force at 2PN order. Because of the helical Killing symmetry we did not include here the 2.5PN radiation-reaction terms; these can be found in Eqs.~(7.6) of \cite{BFP98}. 

In some logarithmic terms at 3PN order $\ell_0$ denotes the arbitrary constant length scale associated with dimensional regularization, which relates the $d$-dimensional gravitational constant $G^{(d)}$ to the usual Newton constant $G$ through \eqref{Geps}. This scale appears conjointly with the numerical combination
\begin{equation}\label{p}
    p \equiv \sqrt{4 \pi} \,e^{C/2}\,,
\end{equation}
where $C = 0.5772\cdots$ is the Euler--Mascheroni constant.\footnote{The number $p$ appears in the expansion when $\varepsilon \rightarrow 0$ of the parameter $k$ defined by Eq.~\eqref{k} as $k = 1 - \varepsilon \ln{p} + \mathcal{O}(\varepsilon^2)$.}

Notice the important feature that the metric in harmonic coordinates involves some \textit{poles} $\propto 1/\varepsilon$ at the 3PN order in the $00$ and $0i$ components, where $\varepsilon$ is related to the spatial dimension $d$ by $d \equiv 3 + \varepsilon$ (see Sec.~\ref{PN}), and formally tends to zero. The results presented in Eqs.~\eqref{g1dr} include the pole part $\sim\varepsilon^{-1}$ and the complete finite part $\sim\varepsilon^0$, and neglect the terms tending to zero when $\varepsilon\rightarrow 0$; for simplicity we do not indicate the remainders $\mathcal{O}(\varepsilon)$. 

However there is an exception to the above rule, in that we have to re-introduce the correction terms $\mathcal{O}(\varepsilon)$ in the \textit{Newtonian} part of the metric. Indeed, when we shall reduce the metric to the center-of-mass frame and then to circular orbits, these corrections will be multiplied by poles at 3PN order, and will contribute \textit{in fine} to the finite part at 3PN order. Such corrections will be necessary only in the $00$ component of the metric, where the 3-dimensional Newtonian potential at the location of the particle 1, namely $V_\mathrm{N}(y_1)=G m_2 / r_{12}$, is to be replaced by its $d$-dimensional version\footnote{Terms $\mathcal{O}(\varepsilon^2)$ are neglected. See \eqref{UNda} and \eqref{aNd} for the exact expressions of the Newtonian potential and acceleration in $d$ dimensions.}
\begin{equation}\label{UNdexp}
    V_\mathrm{N}^{(d)}(y_1) = \frac{G m_2}{r_{12}} \left\{ 1 + \varepsilon \left[ \frac{1}{2} - \ln{\left( \frac{r_{12} \, p}{\ell_0} \right)} \right] \right\} .
\end{equation}

The poles $\propto\varepsilon^{-1}$ in the metric \eqref{g1dr} could be removed by a coordinate transformation and a suitable shift of the two particle's world-lines. This is discussed in Appendix~\ref{HR} where we compute the regularized 3PN metric using the alternative Hadamard regularization. There we show that, modulo some assumptions necessary to overcome the known drawbacks of Hadamard's regularization (\textit{viz} the presence of ambiguities at 3PN order), the regularized metrics in the two regularization schemes are \textit{physically equivalent}, in the sense that they differ by a coordinate transformation plus the additional effect of some shifts of the world-lines of the particles. In particular we find complete agreement with the shifts necessary to link together the 3PN equations of motion computed in Hadamard \cite{BFeom} and dimensional \cite{BDE04} regularizations. However, for the present purpose it is better to leave as they are the poles $\propto\varepsilon^{-1}$ in the metric \eqref{g1dr}, because we are going to compute a gauge invariant quantity, and the poles will ultimately be automatically cancelled in the final result.

Finally we note that the metric depends also on the extra arbitrary constant $r_0$, present in some logarithmic terms of the 00 component of the metric at 3PN order. This constant comes from the IR regularization of the metric at spatial infinity, as discussed in Sec.~\ref{IR}, and it shall also disappear in the final gauge invariant result.

As an important check of the metric \eqref{g1dr} we have verified that it is invariant under a general Lorentz boost, considered in a perturbative 3PN sense. The Lorentz invariance permits checking most of the 2PN terms and also the dynamical 3PN ones. The only terms which are not checked by a 3PN Lorentz boost are the 3PN static ones --- those that do not depend on velocities.

\subsection{The gauge invariant quantity $u^T$}
\label{uT}

To compute the gauge invariant quantity $u^T$ (associated with particle 1 for stationary, circular orbits), we adopt its coordinate form as given by \eqref{ut_defX}, namely
\begin{equation}\label{utPN}
    u^t = \biggl( - g\lab(y_1) \frac{v_1^\alpha v_1^\beta}{c^2}\biggr)^{-1/2}\,,
\end{equation}
and plug into it the 3PN regularized metric explicitly obtained in \eqref{g1dr}. To begin with, this yields the expression of $u^t$ at 3PN order for an arbitrary mass ratio $q=m_1/m_2$, and for a generic non-circular orbit in a general reference frame. 

We then choose the frame of the \textit{center of mass} (CM), which is consistently defined at the 3PN order by the nullity of the 3PN center-of-mass integral of the motion deduced from the 3PN equations of motion \cite{deA.al.01}. We want to express the individual positions $\mathbf{y}_a\equiv\mathbf{y}_a^\mathrm{CM}$ and velocities $\mathbf{v}_a\equiv\mathbf{v}_a^\mathrm{CM}$ (with $a=1,2$ labelling the particles) relatively to the center of mass in terms of the relative position $\mathbf{y}_{12}$ and relative velocity $\mathbf{v}_{12}$. We know how to do this at 3PN order in Hadamard regularization \cite{BI03CM}, and we know that the particle's trajectories in Hadamard regularization differ by a shift of world-lines from those computed with dimensional regularization \cite{BDE04}. So in order to get $\mathbf{y}_a^\mathrm{CM}$ and $\mathbf{v}_a^\mathrm{CM}$ in dimensional regularization we apply directly the shift of world-lines on the known expressions in Hadamard regularization; this is detailed in Appendix~\ref{circ}. 

Having replaced the positions and velocities by their CM expressions $\mathbf{y}_a^\mathrm{CM}[\mathbf{y}_{12},\mathbf{v}_{12}]$ and $\mathbf{v}_a^\mathrm{CM}[\mathbf{y}_{12},\mathbf{v}_{12}]$, the quantity $u^t$ becomes a functional of $\mathbf{y}_{12}$ and $\mathbf{v}_{12}$ which we now reduce to the case of \textit{circular orbits}. This means that $(n_{12}v_{12})=0$ exactly,\footnote{Consistently with the helical Killing symmetry we neglect radiation-reaction effects.} and that the relative orbital velocity squared $v_{12}^2$ takes a specific expression in terms of the relative separation $r_{12}$ or, rather, in terms of the particular dimensionless post-Newtonian parameter defined by
\begin{equation}\label{gamma}
    \gamma \equiv \frac{G\,m}{r_{12} c^2}\,,
\end{equation}
where $m = m_1 + m_2$ is the total mass of the binary.\footnote{Recall that the orbital separation $r_{12}$ is here defined in harmonic coordinates, and differs from the Schwarzschild coordinate distance $r$ used in the SF calculation of Sec.~\ref{overview}.} We find in Appendix~\ref{circ} that the required relation, valid in dimensional regularization, is
\begin{align}\label{vvgammadrtext}
    \frac{v_{12}^2}{c^2} &= \gamma \left\{ 1 + \varepsilon \left[ \frac{3}{2} - \ln{\left( \frac{r_{12} \, p}{\ell_0} \right)} \right] + \left( -3 + \nu \right) \gamma + \left( 6 + \frac{41}{4} \nu + \nu^2 \right) \gamma^2 \right. \\ &+ \left. \left( -10 + \left[ - \frac{2987}{24} + \frac{41}{64} \pi^2 - \frac{11}{\varepsilon} + 44 \ln{\left( \frac{r_{12} \, p}{\ell_0} \right)} \right] \nu + \frac{19}{2} \nu^2 + \nu^3 \right) \gamma^3 + \mathcal{O}(\gamma^{4})\right\} \, ,\nonumber
\end{align}
where $\nu \equiv m_1 m_2 / m^2$ is the symmetric mass ratio, related to the asymmetric mass ratio $q$ by $\nu=q/(1+q)^2$. From now on we assume that $m_1 \leqslant m_2$ to prepare the ground for the small mass ratio case $m_1\ll m_2$ in which $\nu=q+\mathcal{O}(q^2)$. Notice the presence of a pole $\propto 1/\varepsilon$ at the 3PN order in \eqref{vvgammadrtext}, and recall that $\ell_0$ is the dimensional regularization scale, and that $p$ is defined by \eqref{p}. Note also that we have included the $\mathcal{O}(\varepsilon)$ correction in the Newtonian approximation of the expression \eqref{vvgammadrtext}; this is crucial because multiplying that Newtonian term $\mathcal{O}(\varepsilon)$ by a quantity having a pole at 3PN will yield a finite part contribution at 3PN order. The last step of the calculation consists of replacing $\gamma$ by its expansion in powers of the convenient alternative dimensionless gauge invariant PN parameter $x$, directly related to the orbital frequency $\Omega\equiv v_{12}/r_{12}$ by
\begin{equation}\label{x}
    x \equiv \left( \frac{G\,m\,\Omega}{c^3} \right)^{2/3} \,.
\end{equation}
To find $\gamma$ as a power series in $x$ to 3PN order we invert \eqref{vvgammadrtext} and obtain
\begin{align}\label{gammaxdrtext}
    \gamma &= x \left\{ 1 + \varepsilon \left[ -\frac{1}{2} + \frac{1}{3} \ln{\left( \frac{r_{12} \, p}{\ell_0} \right)} \right] + \left( 1 - \frac{\nu}{3} \right) x + \left( 1 - \frac{65}{12} \nu \right) x^2 \right. \\ &+ \left. \left( 1 + \left[ - \frac{251}{72} - \frac{41}{192} \pi^2 + \frac{11}{3 \varepsilon} - \frac{55}{9} \ln{\left( \frac{r_{12} \, p}{\ell_0} \right)} \right] \nu + \frac{229}{36} \nu^2 + \frac{\nu^3}{81} \right) x^3 + \mathcal{O}(x^{4})\right\} \,.\nonumber
\end{align}

When finally replacing $\gamma$ by $x$ we discover most satisfactorily that all the poles $\propto 1/\varepsilon$ cancel out in the final expression for $u^t$, as well as the associated constant $\ell_0$ (and the pure number $p$). Furthermore, the IR constant $r_0$ also disappears from the result when parameterized by the frequency-related parameter $x$. No matter what the mass ratio, our final result for a 3PN, algebraic relationship between $u^T$ (to which $u^t$ now evaluates) and $x$ (or equivalently $\Omega$), is:
\begin{align}\label{uT_PN}
    u^T &= 1 + \left( \frac{3}{4} + \frac{3}{4} \Delta - \frac{\nu}{2} \right) x + \left( \frac{27}{16} + \frac{27}{16} \Delta - \frac{5}{2} \nu - \frac{5}{8} \Delta \, \nu + \frac{\nu^2}{24} \right) x^2 \\ &+ \left( \frac{135}{32} + \frac{135}{32} \Delta - \frac{37}{4} \nu - \frac{67}{16} \Delta \, \nu + \frac{115}{32} \nu^2 + \frac{5}{32} \Delta \, \nu^2 + \frac{\nu^3}{48} \right) x^3 \nonumber \\ &+ \left( \frac{2835}{256} + \frac{2835}{256} \Delta - \left[ \frac{2183}{48} - \frac{41}{64} \pi^2 \right] \nu - \left[ \frac{12199}{384} - \frac{41}{64} \pi^2 \right] \Delta \, \nu \right. \nonumber \\ &\left. + \left[ \frac{17201}{576} - \frac{41}{192} \pi^2 \right] \nu^2 + \frac{795}{128} \Delta \, \nu^2 - \frac{2827}{864} \nu^3 + \frac{25}{1728} \Delta \, \nu^3 + \frac{35}{10368} \nu^4 \right) x^4 + \mathcal{O}(x^5) \,,\nonumber
\end{align}
where we denote $\Delta \equiv (m_2-m_1)/m = \sqrt{1 - 4 \nu}$, so that the test-mass limit of particle 1 corresponds to $\nu\rightarrow 0$. The expression \eqref{uT_PN} is a polynomial in $x$ with coefficients depending only on the symmetric mass ratio $\nu$; it is therefore clearly gauge invariant. While it has been shown in \cite{Det08} (see also Sec.~\ref{method} above) that $u^T$ is gauge invariant at any PN order, in the extreme mass ratio limit $\nu\ll 1$, here we find that it is also gauge invariant for \textit{any} mass ratio up to 3PN order.\footnote{As a test of the initial expression of $u^t$ for a generic orbit in a general frame (i.e. before going to the CM frame), we checked that $\ud u^t / \ud t = 0$ after reduction to circular orbits, as required by the helical symmetry, i.e. neglecting the radiation-reaction.}

\section{Comparison of post-Newtonian and self-force results}
\label{comp}

We now reduce the 3PN expression \eqref{uT_PN} in the small mass ratio regime $q=m_1/m_2 \ll 1$. We express the result in terms of the non-symmetric PN parameter introduced in \eqref{yPN}, which is more suited than $x$ to the small mass ratio limit of particle 1, namely
\begin{equation}\label{y}
    y \equiv \left( \frac{G\,m_2\,\Omega}{c^3} \right)^{2/3} = \frac{G m_2}{R_\Omega c^2}.
\end{equation}
Using $x = y (1 + q)^{2/3}$ and $\nu = q / (1 + q)^2$ we obtain, up to say the quadratic order in $q$,
\begin{align}\label{uT_PN_SF}
    u^T &= 1 + \left( \frac{3}{2} - q + q^2\right) y + \left( \frac{27}{8} - 2 q + 3 q^2\right) y^2 + \left( \frac{135}{16} - 5 q + \frac{97}{8} q^2\right) y^3 \nonumber \\ &+ \left( \frac{2835}{128} + \left[
        -\frac{121}{3} + \frac{41}{32} \pi^2 \right] q + \left[
        \frac{725}{12} - \frac{41}{64} \pi^2 \right] q^2 \right) y^4 + \mathcal{O}(q^3,y^5) \,.
\end{align}
This is to be compared with the result of SF calculations, which take the general form
\begin{equation}\label{utexp}
    u^T = u^T_\mathrm{Schw} + q \, u^T_\mathrm{SF} + q^2 \, u^T_\mathrm{PSF} + \mathcal{O}(q^3) \, ,
\end{equation}
with self-force and post-self-force coefficients $u^T_\mathrm{SF}$ and $u^T_\mathrm{PSF}$ respectively. From \eqref{uT_PN_SF} we thus recover the 3PN expansion of the Schwarzschildean result, i.e.
\begin{equation}\label{utSchw}
    u^T_\mathrm{Schw} = \left( 1 - 3 y \right)^{-1/2} = 1+\frac{3}{2}y+\frac{27}{8}y^2+\frac{135}{16}y^3+\frac{2835}{128}y^4+ \mathcal{O}(y^5)\,.
\end{equation}
Next, we obtain the self-force contribution $u^T_\mathrm{SF}$ up to 3PN order as
\begin{equation}\label{utSF}
    u^T_\mathrm{SF} = - y - 2 y^2 - 5 y^3 + \left(
- \frac{121}{3} + \frac{41}{32} \pi^2 \right) y^4 + \mathcal{O}(y^5) \,.
\end{equation}
The 2PN result is in agreement with \eqref{utSF2PN} as it should. For the much more difficult 3PN coefficient, whose value depends on subtle issues regarding the self-field regularization (see Sec.~\ref{PN}), we thus find
\begin{equation}\label{C3PN}
    \mathcal{C}_\text{3PN} =
- \frac{121}{3} + \frac{41}{32} \pi^2 \,.
\end{equation}
We get also the 3PN expansion of the post-self-force, which could be compared with future SF analyses with second-order black hole perturbations,\footnote{Notice that $u^T_\mathrm{SF} < 0$ and $u^T_\mathrm{PSF} > 0$ (at least up to 3PN order). The effect of the self-force is to reduce the value of $u^T$, while the post-self-force tends to increase it.}
\begin{equation}\label{utpSF}
    u^T_\mathrm{PSF} = y + 3 y^2 + \frac{97}{8} y^3 + \left(
        \frac{725}{12} - \frac{41}{64} \pi^2 \right) y^4 + \mathcal{O}(y^5) \,,
\end{equation}
as well as all higher post-self-force effects up to 3PN order.

\begin{figure}
    \includegraphics[width=10.5cm,angle=-90]{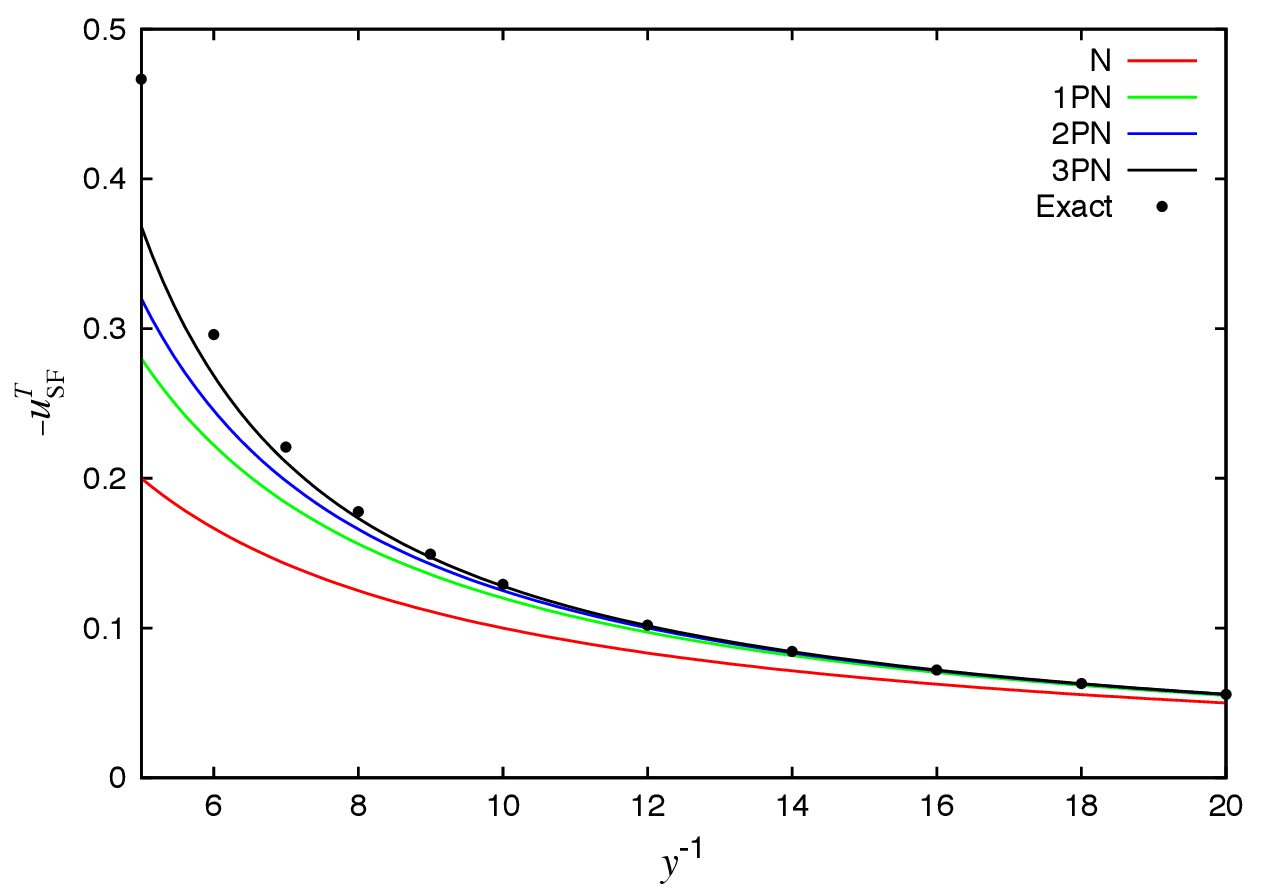}
    \caption{\footnotesize The self-force contribution $u^T_\mathrm{SF}$ to $u^T$ plotted as a function of the gauge invariant variable $y^{-1}$. Note that $y^{-1}$ is equal to $R_\Omega/\m$, an invariant measure of the orbital radius, scaled by the black hole mass $m_2$ [cf. Eq.~\eqref{yPN}]. The ``exact'' numerical points are taken from Ref.~\cite{Det08}.}
    \label{ut_SF}
\end{figure}

Numerically, the 3PN coefficient in the self-force is $\mathcal{C}_\text{3PN} = - 27.6879\cdots$. This shows a remarkable agreement between the post-Newtonian prediction and the result of the numerical SF calculation reported in \eqref{CSF}, namely $\mathcal{C}_\text{3PN}^\text{SF} = - 27.677\pm 0.005$. The two results are consistent at the $2\sigma$ level with five significant digits. This agreement can also be visualized in Fig.~\ref{ut_SF}, where we show the SF contribution $u^T_\mathrm{SF}$ to $u^T$ as a function of $y^{-1}$, as well as the successive Newtonian, 1PN, 2PN and 3PN approximations to $u^T_\mathrm{SF}$. Observe notably the nice convergence of the successive PN approximations toward the exact SF result. The 3PN approximation is roughly $1\%$ accurate up to $y^{-1} = 10$, and roughly $5\%$ accurate up to $y^{-1} = 7$, not very far from the highly relativistic Schwarzschild innermost stable circular orbit (ISCO) for which $y^{-1}_\mathrm{ISCO}=6$.\footnote{See \cite{BS09} for a recent calculation of the shift of the Schwarzschild ISCO induced by the conservative part of the self-force.}

This successful comparison between SF and PN calculations confirms the soundness of both approximations in describing compact binary systems.  In the post-Newtonian calculation, this encompasses the post-Newtonian expansion as applied to the binary equations of motion \cite{JaraS98,BFeom}, and includes the treatment of the issues associated with the UV divergencies using dimensional regularization \cite{DJSdim,BDE04}.  In addition, the IR divergences, too, in the PN calculation (see Sec.~\ref{IR}) are seen to be correctly treated, since their effects vanish in the final result \eqref{uT_PN}.  In the perturbative self-force calculation embodied in \eqref{utSF2PN}--\eqref{CSF}, this includes the delicate handling of gauge and the numerically taxing split of the metric near the particle into singular and regular pieces following the prescriptions in \cite{DW03}.  In this light, it would be interesting to address the opposite question, namely that of estimating the accuracy of the black hole perturbation formalism by comparing several truncated self-force series to the ``exact'' PN result in the slow motion limit. This would require at least a second-order perturbative SF calculation.

\begin{figure}
    \includegraphics[width=15cm,angle=0]{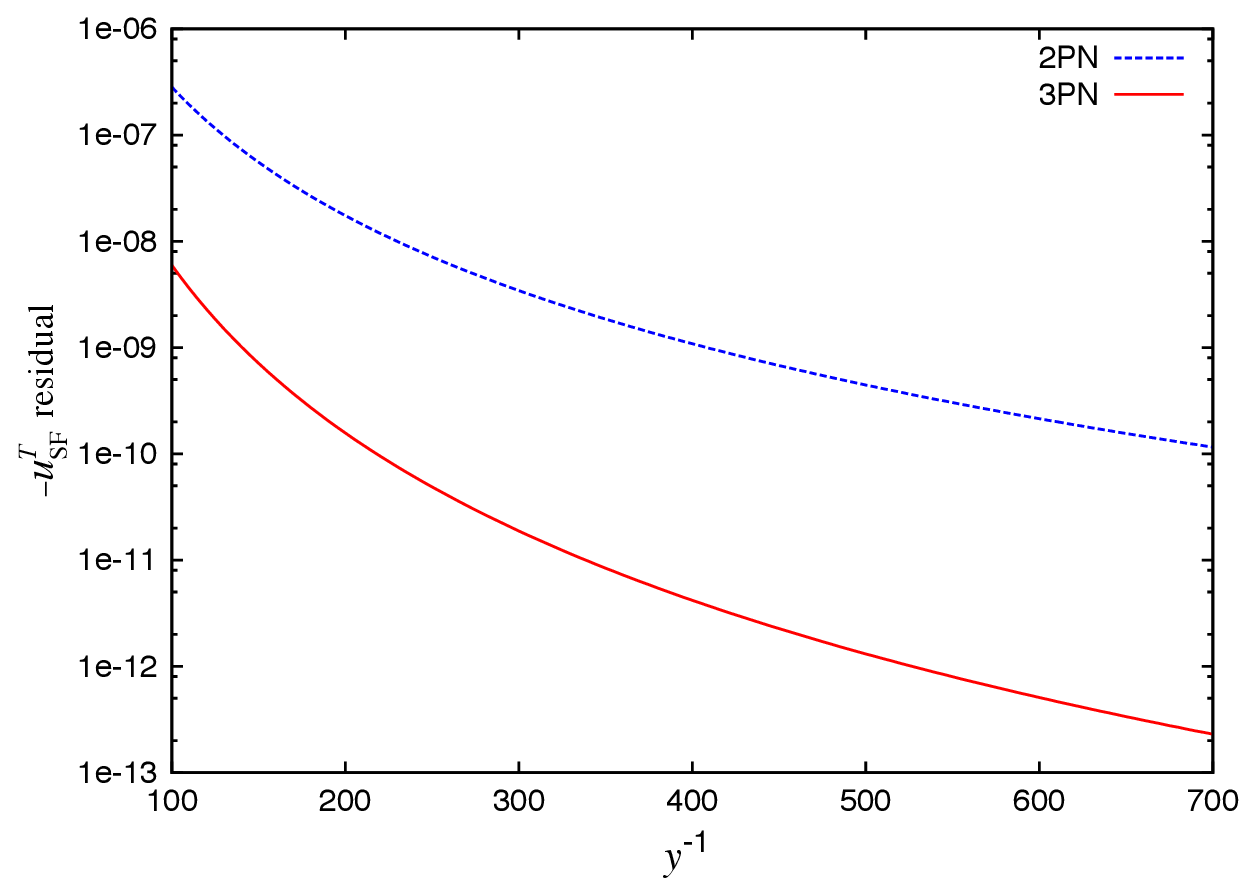}
    \caption{\footnotesize Numerically derived residuals, i.e., after removal of the 2PN and 3PN self-force contributions to $u^T_\mathrm{SF}$, plotted as a function of the gauge invariant variable $y^{-1}$. Compare with scales in Fig.~\ref{ut_SF}.  Note that $y^{-1}$ is equal to $R_\Omega/\m$, an invariant measure of the orbital radius, scaled by the black hole mass $m_2$ [cf. \eqref{yPN}].}
    \label{ut_3PN}
\end{figure}

Our post-Newtonian calculation contains additional results which have not been used in this paper.  For example, it already contains some of these higher-order self-force terms, as evidenced by \eqref{utpSF}.  Similarly, our numerical self-force calculation actually contains much more information than is indicated by the numerical coefficient we give in \eqref{CSF}. This is most simply illustrated in Fig.~\ref{ut_3PN}, where we show, over the large-$R_\Omega$ range used for our numerical fitting, the full 2PN and 3PN residuals, that is, the residuals after the known 2PN and (now) 3PN terms have been subtracted from our numerical data.  In fact, we have gone to considerable lengths to ensure that we would have high quality numerical data to work with here. The smooth curve of 3PN residuals, several orders of magnitude below the 2PN curve, is a testament to this data quality and represents the starting point for an investigation which more adequately explains the appropriate higher order PN nature of our numerical data; especially the presence of logarithmic terms in higher PN approximations. The pressing need for this explanation is strong motivation for further work \cite{BDLW09b}.

\section*{Acknowledgements}

SD and BFW acknowledge support through grants PHY-0555484 and PHY-0855503 from the National Science Foundation.  LB thanks the University of Florida for a visit supported by the Programme International de Coop\'eration Scientifique (CNRS--PICS).  All authors acknowledge the 2008 Summer School on Mass and Motion, organized by A.~Spallicci and supported by the University of Orl\'eans and the CNRS, through which we experienced an extensive opportunity to understand each other's perspective and make rapid progress on this work.

\appendix

\section{Relation to Hadamard regularization}
\label{HR}

As an important check of the DR calculation of $u^T$, we have also performed the complete calculation using the alternative Hadamard regularization (HR), in the variant proposed in \cite{BFeom} and called the ``extended Hadamard regularization''. The HR is essentially based on the Hadamard partie finie integral \eqref{partiefinie}. Unfortunately we know that the HR, even in the extended variant, is not entirely satisfying and cannot give a completely consistent picture at the 3PN order \cite{JaraS98,BFeom}. In particular it yields some ambiguities in the form of a small number of arbitrary parameters which cannot be computed within this regularization. However, the HR gives the correct answer provided that the ambiguity parameters are specified by some external arguments, or fixed by comparison with some non-ambiguous calculations. Then it becomes a non trivial check to show that it is possible to adjust a few HR ambiguity parameters so that the \textit{complete} result, which is generally made of many more terms, agrees with the result of DR.

The ambiguity parameters in HR come from the unknown relations between two sets of arbitrary length scales denoted $s_a$ and $r_a'$ (where $a$ labels the particles). Here the scales $s_a$ are introduced into the Hadamard partie finie \cite{Hadamard,Schwartz} of Poisson integrals with singular sources, when computed at any field point $\mathbf{x}'$ different from the singularities $\mathbf{y}_a$. The $s_a$'s appear when the Poisson integrals develop logarithmic divergences on the singular points (i.e. when the source point $\mathbf{x}$ over which one integrates equals $\mathbf{y}_a$). The other scales $r_a'$ come from the singular limit of the Poisson potential when the \textit{field} point $\mathbf{x}'$ itself tends toward the singularity $\mathbf{y}_a$; hence we have in fact $r_a'=\vert\mathbf{x}'-\mathbf{y}_a\vert$ which shows up in the form of some ``constant'' $\ln r_a'$ which is formally infinite. It was shown that the relation between the scales $s_a$ and $r_a'$ must involve the masses $m_a$ (and $m\equiv\sum_bm_b$), and is necessarily of the type \cite{BFeom}
\begin{equation}\label{lograsa}
    \ln\left(\frac{r_a'}{s_a}\right)=\alpha+\beta\frac{m}{m_a}\,.
\end{equation}
Here $\alpha$ and $\beta$ denote some purely numerical constants and are called \textit{ambiguity parameters}. After imposing the link \eqref{lograsa} to get rid of the scales $s_a$, it was shown that the remaining scales $r_a'$ are gauge constants which can be removed by a change of gauge.\footnote{Hence the fact that $\ln r_a'$ is actually ``infinite'' does not really matter.} In the case of the 3PN equations of motion (EOM), and for the extended variant of HR, it was found that the correct values are \cite{BDE04}\footnote{More precisely, $\alpha_\mathrm{EOM}$ was determined by requiring that the equations of motion should derive from a Lagrangian formulation, but $\beta_\mathrm{EOM}$ (which was denoted $\lambda$ in Refs.~\cite{BFeom,BDE04}) had to wait until its value was fixed by DR.}
\begin{equation}\label{alphabetaeom}
    \alpha_\mathrm{EOM} = \frac{159}{308}\quad\text{and}\quad\beta_\mathrm{EOM} = - \frac{1987}{3080}\,.
\end{equation}
In the case of the 3PN mass quadrupole (MQ) moment needed to compute the 3PN radiation field, the values using the same extended HR turned out to be \cite{BI04mult,BDEI04}\footnote{The coefficients $\alpha_\mathrm{MQ}$ and $\beta_\mathrm{MQ}$ were respectively denoted $\xi$ and $\kappa$ in Refs.~\cite{BI04mult,BDEI04}.}
\begin{equation}\label{alphabetamq}
    \alpha_\mathrm{MQ}=-\frac{9451}{9240}\quad\text{and}\quad\beta_\mathrm{MQ}=0\,.
\end{equation}
The fact that the MQ values are different from the EOM values already shows that the HR is not satisfying: Two different computations give inconsistent determinations of the ambiguity parameters. For the present computation of the quantity $u^T$ and comparison with SF calculations, we have shown that the extended variant of HR reproduces exactly the result of DR [i.e. \eqref{uT_PN} above] \textit{if and only if} we have the still different values
\begin{equation}\label{alphabeta}
    \alpha_\mathrm{SF} =-\frac{65}{154}\quad\text{and}\quad\beta_\mathrm{SF}=\frac{129}{440}\,.
\end{equation}
Although this result shows again that the Hadamard regularization is not consistent at 3PN order, we argue that it constitutes a powerful check of our calculation, because for the complete agreement we have to adjust no more than two unknown coefficients. In particular we find that the HR constants $r_a'$, which remain after imposing the relation \eqref{lograsa}, cancel out properly in the final result. The HR calculation is also interesting because it corresponds to a different harmonic coordinate system and a different definition of the particle's world-lines. Since we have the regularized 3PN metric in both HR and DR, we can now study in detail their difference --- adopting the values \eqref{alphabeta} in the HR scheme.

We shall find that the two metrics differ by an infinitesimal 3PN coordinate transformation in the ``bulk'', i.e. outside the particle's world-lines, and also by an intrinsic \textit{shift} of these world-lines. In particular we shall recover the total shift obtained at the level of the equations of motion in Ref.~\cite{BDE04}, but shall prove that this shift is made of the latter intrinsic shift, plus the shift induced by the coordinate transformation. Let the coordinate transformation between the two metrics be $\delta x^\alpha = \epsilon^\alpha(x)$, where $\epsilon^0=\mathcal{O}(c^{-7})$ and $\epsilon^i=\mathcal{O}(c^{-6})$ as appropriate to 3PN order. The transformation of the bulk metric is (for simplicity we omit the PN remainders)
\begin{subequations}
    \begin{align}
        \delta_\epsilon g_{00} &= - 2 \partial_0 \epsilon_0 - \epsilon^i \partial_i g_{00} - \epsilon^i(y_1) \frac{\partial g_{00}}{\partial y_1^i} - \epsilon^i(y_2) \frac{\partial g_{00}}{\partial y_2^i} \, , \label{g001drhr} \\
        \delta_\epsilon g_{0i} &= - 2 \partial_{(0} \epsilon_{i)} \, , \\
        \delta_\epsilon g_{ij} &= - 2 \partial_{(i} \epsilon_{j)} \, . \label{gij1drhr}
    \end{align}
\end{subequations}
The terms involving partial derivatives with respect to the source points $\mathbf{y}_a$ take into account the shifts of the trajectories $\bm{\epsilon}(y_a)$ through a modification of the source dependence of the metric $g\lab(x;y_a,v_a)$; we use the notation $y_a\equiv(c t,\mathbf{y}_a)$ and $v_a\equiv(c,\mathbf{v}_a)$. Since $\epsilon^\alpha$ is of order 3PN, the $g_{00}$'s in the RHS of \eqref{g001drhr} are simply Newtonian. At the point 1 we get\footnote{Note that our too compact notation $g_{00}(y_1)$ stands in fact for $g_{00}(y_1;y_a,v_a)$; thus we have used the obvious relations $\partial[g_{00}(y_1)]/\partial y_1^i=(\partial_i g_{00})(y_1)+(\partial g_{00}/\partial y_1^i)(y_1)$ and $\partial[g_{00}(y_1)]/\partial y_2^i=(\partial g_{00}/\partial y_2^i)(y_1)$.}
\begin{subequations}\label{deltaepsilon1}
    \begin{align}
        \delta_\epsilon g_{00}(y_1) &= - 2 \partial_0 \epsilon_0(y_1) - \epsilon^i(y_1) \frac{\partial}{\partial y_1^i}\bigl[g_{00}(y_1)\bigr] - \epsilon^i(y_2) \frac{\partial}{\partial y_2^i}\bigl[g_{00}(y_1)\bigr] \, , \label{deltaepsg001} \\
        \delta_\epsilon g_{0i}(y_1) &= - 2 \partial_{(0} \epsilon_{i)}(y_1) \, , \\
        \delta_\epsilon g_{ij}(y_1) &= - 2 \partial_{(i} \epsilon_{j)}(y_1) \, .\label{deltaepsgij1}
    \end{align}
\end{subequations}
Now we have found that in order to relate the two metrics one must additionally perform a shift $\bm{\kappa}_a$ of the particle's trajectories at the 3PN order, i.e. $\bm{\kappa}_a=\mathcal{O}(c^{-6})$. Such shift will be ``\textit{intrinsic}'' in the sense that it will not be induced by any coordinate transformation of the bulk metric. It yields the additional change of the metric components evaluated at point $y_1$:
\begin{equation}\label{deltakappag001}
        \delta_\kappa g_{00}(y_1) = - \kappa_1^i \frac{\partial g_{00}}{\partial y_1^i}(y_1) - \kappa_2^i \frac{\partial g_{00}}{\partial y_2^i}(y_1) \, ,
\end{equation}
while there is no change in the other components at that order, i.e. $\delta_\kappa g_{0i}(y_1) = \delta_\kappa g_{ij}(y_1)= 0$. Our final result is that the two regularized metrics at point 1 are related through
\begin{equation}\label{DRHR}
        g\lab^\mathrm{DR}(y_1) = g\lab^\mathrm{HR}(y_1) + \delta_\epsilon g\lab(y_1) + \delta_\kappa g\lab(y_1) \,.
\end{equation}
This relation is a functional equality relating the two metric functionals at point 1, whose coordinates $y_1\ua$ can be seen as dummy variables.

An important check of our finding \eqref{DRHR} is that it contains the previous result derived at the level of the 3PN equations of motion (not considering the bulk metric), namely that the DR and HR equations of motion merely differ by some shifts $\bm{\xi}_a$ of the particle's world-lines. This result established the physical equivalence of HR and DR at the level of the equations of motion \cite{BDE04}. Indeed, we discover that the \textit{total} shift found here, which is made up of the shift $\bm{\epsilon}(y_a)$ induced by the coordinate transformation plus the intrinsic shift $\bm{\kappa}_a$, is precisely equal to the shift of the world-lines of the particles found in \cite{BDE04}. Thus,
\begin{equation}\label{xiadef}
        \bm{\xi}_a = \bm{\epsilon}(y_a) + \bm{\kappa}_a \,.
\end{equation}
So we have proved that $\bm{\xi}_a$ is partly intrinsic and partly induced by a coordinate transformation of the bulk metric. Since there is no physics involved in a coordinate transformation, it can be argued that only the intrinsic part of the shift $\bm{\kappa}_a$ corresponds to the physical process of renormalization which was performed in \cite{BDE04}.

We give now the explicit expressions. The coordinate transformation at any field point $x = (ct, \mathbf{x})$ reads
\begin{subequations}\label{eps}
    \begin{align}
        \epsilon_0(x) & = \frac{7}{5} \frac{G^3 m_1 m_2^2}{c^7} \, \partial_t \left( \frac{k}{r_1^{1+\varepsilon}} \right) + \frac{12}{5} \frac{G^3 m_1 m_2^2}{c^7} \, v_{12}^i \partial_i \left(  \frac{k}{r_1^{1+\varepsilon}} \left[ \frac{1}{\varepsilon} - 2 \ln{\left( \frac{r'_2 \, p}{\ell_0} \right)} - \frac{301}{180} \right]\right) \nonumber \\ &+ 1 \leftrightarrow 2 + f_1(t) \, , \label{eps0} \\
        \epsilon_i(x) & = \frac{7}{5} \frac{G^3 m_1 m_2^2}{c^6} \, \partial_i \left( \frac{k}{r_1^{1+\varepsilon}} \right) + 1 \leftrightarrow 2 \,. \label{epsi}
    \end{align}
\end{subequations}
Remind that the DR-related quantites $p$, $\ell_0$ and $k$ are defined by Eqs.~\eqref{p}, \eqref{Geps} and \eqref{k} respectively, while $r_2'$ is an HR constant of particle 2. The symbol $1 \leftrightarrow 2$ means adding the previous expression [i.e. \textit{excluding} the term $f_1(t)$ in \eqref{eps0}], but with all particle labels exchanged (thus $r_2'$ would be changed to $r_1'$). Note that $r_1'$ and $r_2'$ are considered as true constants in \eqref{eps}. The function $f_1(t)$ is given by
\begin{equation}\label{f1}
        f_1(t) = \frac{91}{15} \frac{G^3 m_1^2 m_2}{c^7} \,\frac{k(1+\varepsilon)}{r_{12}^{2+\varepsilon}}\, (n_{12}v_{12}) \left[ \frac{1}{\varepsilon} - 2 \ln{\left( \frac{r'_1 \, p}{\ell_0} \right)} + \frac{1642}{1365} \right] .
\end{equation}
Notice that because of the presence of $f_1(t)$ in \eqref{eps0}, the time component of the gauge vector $\epsilon_0(x)$ is not symmetric by exchange $1 \leftrightarrow 2$. This coordinate transformation satisfies the harmonic gauge condition $\Box \epsilon^\alpha = 0$ in $d$ dimensions at the 3PN accuracy. We note also that the time component $\epsilon_0$ of the coordinate transformation has a pole part $\propto \varepsilon^{-1}$, as well as a pole-independent part, but that the space component $\epsilon_i$ is pole-free.

Beware that strictly speaking \eqref{eps} is not the coordinate transformation between the HR metric $g^\text{HR}\lab(x)$ and the DR metric $g^\text{DR}\lab(x)$ in the bulk. It is solely the restriction $\epsilon^\alpha(y_1)$ of this gauge transformation at the location of particle $1$ that correctly relates the two regularized metrics $g^\text{HR}\lab(y_1)$ and $g^\text{DR}\lab(y_1)$ at that location. Indeed, if the gauge transformation \eqref{eps} was to be valid in the bulk, it would induce poles in the $00$ and $0i$ components of $g^\text{DR}\lab(x)$. But it was shown in \cite{BDE04}, based on diagrammatic arguments, that such poles, at 3PN order, can only be present in the $00$ component $g^\text{DR}_{00}(x)$ of the DR metric. However, the restriction $\epsilon^\alpha(y_1)$ at the location of particle $1$ of the pole-free gauge transformation in the bulk does generate poles in the $0i$ component of the DR metric at $y_1$ [see Eq.~\eqref{g0i1dr}].

The shift induced by this coordinate transformation is pole-free, and we immediately get from \eqref{epsi} [up to a correction $\mathcal{O}(\varepsilon)$]
\begin{equation}
    \bm{\epsilon}(y_a) = \frac{7}{5} \frac{G^2 m_a^2}{c^6} \, \mathbf{a}^\text{N}_a \, ,
\end{equation}
where $\mathbf{a}^\text{N}_a$ is the $d$-dimensional Newtonian acceleration of body $a$ given by \eqref{aNd} below. Because $\bm{\epsilon}(y_a)$ does not contain any pole, we observe from \eqref{deltaepsgij1} that the spatial part of the regularized metric will be free of poles at 3PN order as well [cf. Eq.~\eqref{gij1dr}]. Next we find that the additional shift $\bm{\kappa}_a$ does contain a pole, and explicitly reads
\begin{equation}\label{kappaa}
    \bm{\kappa}_a = \frac{11}{3} \frac{G^2 m_a^2}{c^6} \left[ \frac{1}{\varepsilon} - 2 \ln{\left( \frac{r'_a \, p}{\ell_0} \right)} - \frac{\text{183}}{\text{308}} \right] \mathbf{a}^\text{N}_a \, ,
\end{equation}
so that the total shift as defined by \eqref{xiadef} is given by
\begin{equation}\label{xia}
    \bm{\xi}_a = \frac{11}{3} \frac{G^2 m_a^2}{c^6} \left[ \frac{1}{\varepsilon} - 2 \ln{\left( \frac{r'_a \, p}{\ell_0} \right)} - \frac{\text{327}}{\text{1540}} \right] \mathbf{a}^\text{N}_a \, ,
\end{equation}
in perfect agreement with the result of \cite{BDE04}. 

For completeness we now give the result for the difference between the two regularized metrics, $\delta g\lab(y_1) \equiv g\lab^\mathrm{DR}(y_1) - g\lab^\mathrm{HR}(y_1)$. Combining \eqref{deltaepsilon1}--\eqref{DRHR} with \eqref{eps}--\eqref{kappaa} we get
\begin{subequations}
    \begin{align}
        \delta g_{00}(y_1) &= \frac{G^3 m_1^2 m_2}{c^8 r_{12}^3} \, \biggl\{ \frac{2308}{25} (n_{12}v_1)^2 - \frac{4738}{25} (n_{12}v_1)(n_{12}v_2) + \frac{444}{5} (n_{12}v_2)^2 \nonumber \\ &- \frac{6014}{225} v_1^2 + \frac{12754}{225} (v_1 v_2) - \frac{1222}{45} v_2^2 + \left[ \frac{1}{\varepsilon} - 3 \ln{\left( \frac{r_{12} \, p}{\ell_0} \right)} + 2 \ln{\left( \frac{r_{12}}{r'_1} \right)} \right] \times \nonumber \\ &\biggl( \frac{182}{5} \left[ (n_{12}v_1)^2 - \frac{1}{3} v_1^2 \right] - \frac{292}{5} \left[ (n_{12}v_1)(n_{12}v_2) - \frac{1}{3} (v_1 v_2) \right] + 22 \left[ (n_{12}v_2)^2 - \frac{1}{3} v_2^2 \right] \biggr) \nonumber \\ &+ \frac{22}{3} \frac{G m}{r_{12}} \left[ \frac{1717}{330} + \frac{1}{\varepsilon} - 4 \ln{\left( \frac{r_{12} \, p}{\ell_0} \right)} + 2 \ln{\left( \frac{r_{12}}{r'_1} \right)} \right] \biggr\} \nonumber \\ &- \frac{22}{3} \frac{G^4 m_1 m_2^3}{c^8 r_{12}^4} \left[ \frac{4293}{1540} + \frac{1}{\varepsilon} - 4 \ln{\left( \frac{r_{12} \, p}{\ell_0} \right)} + 2 \ln{\left( \frac{r_{12}}{r'_2} \right)} \right] , \\
        \delta g_{0i}(y_1) &= \frac{G^3 m_1^2 m_2}{c^7 r_{12}^3} \biggl\{ \frac{61}{25} (n_{12}v_1) n_{12}^i + \frac{149}{25} (n_{12}v_2) n_{12}^i - \frac{121}{75} v_1^i - \frac{89}{75} v_2^i \nonumber \\ &+\frac{12}{5} \left[ v_{12}^i - 3 (n_{12} v_{12}) n_{12}^i \right] \left[ \frac{1}{\varepsilon} - 3 \ln{\left( \frac{r_{12} \, p}{\ell_0} \right)} + 2 \ln{\left( \frac{r_{12}}{r'_1} \right)} \right] \biggr\} \, , \\
        \delta g_{ij}(y_1) &= \frac{14}{5} \frac{G^3 m_1^2 m_2}{c^6 r_{12}^3} \left( \delta^{ij} - 3 n_{12}^i n_{12}^j \right) .
    \end{align}
\end{subequations}
The end result for the Hadamard regularized 3PN metric, $g\lab^\mathrm{HR}(y_1)$, then follows from combining the previous difference with the explicit expression \eqref{g1dr} for the DR metric. One can check while performing the sum that all poles $ \propto 1/\varepsilon$ and the associated $\ell_0$-dependent logarithmic terms cancel out, so that the HR result only depends on the UV gauge constants $r'_a$ and also, of course, on the IR regularization constant $r_0$.

Note that the DR metric \eqref{DRHR} is really the metric experienced by the particle in $d = 3 + \varepsilon$ dimensions. It is thus very important to include in that metric all corrections of order $\varepsilon$ which could yield finite contributions after multiplication by quantities involving poles. As already mentioned, for the problem of computing $u^T$ for circular orbits we have to write the Newtonian part of the 00 component of the metric as $g_{00}^\text{DR} = - 1 + 2 V^{(d)}_\mathrm{N} / c^{2} + \mathcal{O}(c^{-4})$, where the Newtonian potential satisfying the $d$-dimensional Poisson equation $\Delta V_\text{N}^{(d)} = - 4 \pi G^{(d)} \sigma_\text{N}$ with Newtonian source density $\sigma_\text{N} = \frac{2(d-2)}{d-1} \sum_a m_a \delta_a^{(d)}$ is given by
\begin{equation}\label{UNd}
    V^{(d)}_\mathrm{N}(x) = \frac{2(d-2)}{d-1} k \sum_a \frac{G^{(d)} m_a}{r_a^{d-2}} \, ,
\end{equation}
with DR value at point $a$ [see also \eqref{UNdexp}]
\begin{equation}\label{UNda}
    V^{(d)}_\mathrm{N}(y_a) = \frac{2(d-2)}{d-1} k \sum_{b \neq a} \frac{G^{(d)} m_b}{r_{ab}^{d-2}} \, .
\end{equation}
In the same vein the Newtonian acceleration in \eqref{kappaa}--\eqref{xia} should read
\begin{equation}\label{aNd}
    \mathbf{a}_a^\mathrm{N} = \bm{\nabla} V^{(d)}_\mathrm{N}(y_a) = - \frac{2(d-2)^2}{d-1} k \sum_{b \neq a} \frac{G^{(d)} m_b}{r_{ab}^{d-1}} \, \mathbf{n}_{ab} \, .
\end{equation}

\section{Circular orbits in $d$ dimensions}
\label{circ}

In this Appendix we describe our way to reduce a general $d$-dimensional expression such as the regularized metric \eqref{g1dr} ---valid for arbitrary binary orbits and in a general frame (in harmonic coordinates)--- to the center-of-mass (CM) frame and then to circular orbits. The relevant formulas to do so have been worked out at 3PN order within HR (see \cite{Bliving} for more details), and we need here the corresponding formulas valid in DR. Basically we shall rely on the HR results and apply to them the known shifts of the particle's world-lines to deduce the corresponding DR results.

The 3PN equations of motion of compact binaries using HR turned out to depend on one, and only one, ambiguity parameter called $\lambda$ (denoted $\beta_\text{EOM}$ in Appendix~\ref{HR}) \cite{BFeom}, and to be physically equivalent to the DR equations of motion if and only if $\lambda = - \frac{1987}{3080}$ \cite{BDE04}.\footnote{This result is equivalent to the one of Ref.~\cite{DJSdim}; see also \cite{itoh1,itoh2} for an alternative, ambiguity-free derivation of the 3PN equations of motion.} This means that the difference between the DR and HR accelerations of body $1$ (say) is exclusively due to a shift of the world-lines of the particles $\mathbf{y}_a \rightarrow \mathbf{y}_a + \bm{\xi}_a$ through
\begin{equation}\label{eom_dr}
    \mathbf{a}_1^\text{DR} = \mathbf{a}_1^\text{HR} \vert_{\lambda = - \frac{1987}{3080}} + \delta_{\xi} \mathbf{a}_1 \,.
\end{equation}
The explicit $\lambda$-dependent expression of the 3PN-accurate acceleration $\mathbf{a}_1^\text{HR}$ can be found in Eq.~(7.16) of \cite{BFeom}. The effect of the shifts $\bm{\xi}_a$ on the acceleration of body $1$ is
\begin{equation}\label{deltaa}
    \delta_{\xi} \mathbf{a}_1 = \ddot{\bm{\xi}}_1 - \xi_{12}^i \, \frac{\partial\mathbf{a}_1^\mathrm{N}}{\partial y_1^i} + \mathcal{O}(c^{-8}) \, ,
\end{equation}
where $\xi_{12}^i\equiv \xi_1^i - \xi_2^i$, and the dot stands for a derivative with respect to coordinate time $t$. The shift $\bm{\xi}_a$ has been given in \eqref{xia} above; recall the presence therein of a pole $\propto\varepsilon^{-1}$. To be consistent one needs to include in the Newtonian acceleration $\mathbf{a}_1^\mathrm{N}$ the corrections of order $\varepsilon$, and the correct expression to do so is given by \eqref{aNd}. 

By definition, the CM frame is such that the center of mass position $\mathbf{G}$ vanishes. Within HR, we have $\mathbf{G}^\text{HR} = \mathbf{0}$ when the individual positions of the particles $\mathbf{y}_a$ are given as some functionals of the relative position $\mathbf{y}_{12}$ and velocity $\mathbf{v}_{12}$ according to $\mathbf{y}_a = \mathbf{y}_a^\text{HR}[\mathbf{y}_{12},\mathbf{v}_{12}]$. The explicit expression of the functionals $\mathbf{y}_a^\text{HR}[\mathbf{y}_{12},\mathbf{v}_{12}]$ up to 3PN order can be found in Eqs.~(3.6)--(3.7) of \cite{BI03CM}. Similarly, within DR we shall have $\mathbf{G}^\text{DR} = \mathbf{0}$ when the individual positions of the particles are related to the relative position and velocity according to some new functional relations
\begin{equation}\label{yaDR}
    \mathbf{y}_a = \mathbf{y}_a^\text{DR}[\mathbf{y}_{12},\mathbf{v}_{12}]\,,
\end{equation}
which we want to determine. Now, by the effect of the shifts of the world-lines the expression of the center of mass position in DR will be different from that in HR, and be given by $\mathbf{G}^\text{DR} = \mathbf{G}^\text{HR} + \delta_{\xi} \mathbf{G}$ where
\begin{equation}\label{dxiG}
    \delta_{\xi} \mathbf{G} = - m_1 \bm{\xi}_1 - m_2 \bm{\xi}_2 + \mathcal{O}(c^{-8}) \, .
\end{equation}
Therefore we find that the DR functionals $\mathbf{y}_a^\text{DR}[\mathbf{y}_{12},\mathbf{v}_{12}]$ are related to the HR functionals $\mathbf{y}_a^\text{HR}[\mathbf{y}_{12},\mathbf{v}_{12}]$ through $\mathbf{y}_a^\text{DR} = \mathbf{y}_a^\text{HR} + \delta_{\xi} \mathbf{y}_a$,\footnote{Note that we mean by this a \textit{functional} equality, valid for any dummy variables $\mathbf{y}_{12}$ and $\mathbf{v}_{12}$.} with the same shift for both particles given by
\begin{equation}\label{yrel}
    \delta_{\xi} \mathbf{y}_a = \frac{m_1}{m} \,\bm{\xi}_1 + \frac{m_2}{m} \,\bm{\xi}_2 + \mathcal{O}(c^{-8})\,.
\end{equation}
The DR expressions \eqref{yaDR} are thus easily determined from the HR results.

Next, from the DR equations of motion \eqref{eom_dr}--\eqref{deltaa} in a general frame, we go to the CM frame by replacing the individual positions and velocities by the relative ones according to $\mathbf{y}_a = \mathbf{y}_a^\text{DR}[\mathbf{y}_{12},\mathbf{v}_{12}]$ and also $\mathbf{v}_a = \dot{\mathbf{y}}_a^\text{DR}[\mathbf{y}_{12},\mathbf{v}_{12}]$. Turning off the well-known 2.5PN radiation-reaction terms, and restricting the result to circular orbits [thus $(y_{12} v_{12})=0$], we get the relative acceleration of the binary within DR in the form $\mathbf{a}_{12}^\text{DR} = - \Omega^2\,\mathbf{y}_{12}$, where the orbital frequency $\Omega$ can then be computed iteratively as an expansion in powers of the PN parameter $\gamma \equiv G m/(r_{12} c^2)$, with $r_{12} = \vert \mathbf{y}_{12} \vert$. To 3PN order we find
\begin{align}\label{omegadr}
    \Omega^2 &= \frac{G m}{r_{12}^3} \left\{ 1 + \varepsilon \left[ \frac{3}{2} - \ln{\left( \frac{r_{12} \, p}{\ell_0} \right)} \right] + \left( -3 + \nu \right) \gamma + \left( 6 + \frac{41}{4} \nu + \nu^2 \right) \gamma^2 \right. \\ &+ \left. \left( -10 + \left[ - \frac{2987}{24} + \frac{41}{64} \pi^2 - \frac{11}{\varepsilon} + 44 \ln{\left( \frac{r_{12} \, p}{\ell_0} \right)} \right] \nu + \frac{19}{2} \nu^2 + \nu^3 \right) \gamma^3 + \mathcal{O}(c^{-8})\right\}  \,.\nonumber
\end{align}
As in the Hadamard case, we invert this relation to express $\gamma$ as a PN series in powers of the gauge invariant parameter $x \equiv ( G m \Omega/c^3 )^{2/3}$, with result
\begin{align}\label{gammaxdr}
    \gamma &= x \left\{ 1 + \varepsilon \left[ -\frac{1}{2} + \frac{1}{3} \ln{\left( \frac{r_{12} \, p}{\ell_0} \right)} \right] + \left( 1 - \frac{\nu}{3} \right) x + \left( 1 - \frac{65}{12} \nu \right) x^2 \right. \\ &+ \left. \left( 1 + \left[ - \frac{251}{72} - \frac{41}{192} \pi^2 + \frac{11}{3 \varepsilon} - \frac{55}{9} \ln{\left( \frac{r_{12} \, p}{\ell_0} \right)} \right] \nu + \frac{229}{36} \nu^2 + \frac{\nu^3}{81} \right) x^3 + \mathcal{O}(c^{-8})\right\}  \,.\nonumber
\end{align}
The equations \eqref{omegadr}--\eqref{gammaxdr} are the DR equivalent of Eqs.~(188) and (191) in \cite{Bliving}, which are valid in the coordinate system used in the HR case. Of course the results coincide up to 2PN order as they should. Notice however that we kept the terms proportional to $\varepsilon$ in the Newtonian terms of \eqref{omegadr}--\eqref{gammaxdr}, because in the process of computing $u^T$ the Newtonian terms will get multiplied by some poles $\varepsilon^{-1}$ occurring at 3PN order, and these corrections will contribute to the final result. We are done for the results necessary for the computation of $u^T$ for circular orbits as reported in Sec.~\ref{PNres}. 

As a useful check, we compute the total energy of the binary for circular orbits within DR, making use of Eqs.~\eqref{yrel}--\eqref{gammaxdr}. For a generic orbit and in a general frame, the DR energy functional is related to the HR one through $E^\text{DR} = E^\text{HR} + \delta_{\xi} E$. The effect of the shifts $\bm{\xi}_a$ on the energy explicitly reads
\begin{equation}\label{deltaxiE}
    \delta_{\xi} E = - m_1 \,v^i_1 \,\dot{\xi}^i_1 - m_2 \,v^i_2 \,\dot{\xi}^i_2 + \xi_{12}^i \, \frac{\partial U^{(d)}_\mathrm{N}}{\partial y_{12}^i} + \mathcal{O}(c^{-8}) \, ,
\end{equation}
where the Newtonian gravitational potential energy in $d$ dimensions is
\begin{equation}\label{VNd}
    U^{(d)}_\mathrm{N} = \frac{2(d-2)}{d-1} k \, \frac{G^{(d)} m_1 m_2}{r_{12}^{d-2}} \, .
\end{equation}
At this stage, we use the expression of the total energy $E^\text{HR}$ as computed within HR, and given e.g. by Eq.~(170) of \cite{Bliving}, and add to it the term $\delta_{\xi} E$ defined by \eqref{deltaxiE}--\eqref{VNd}. Our first check is that this 3PN-accurate energy $E^\text{DR}$ for a generic orbit in a general frame within DR is conserved, i.e. $\dot{E}^\text{DR}=0$ when neglecting the 2.5PN radiation-reaction terms. This requires consistently order reducing the result, i.e. replacing the accelerations in the time derivative of $E^\text{DR}$ using the DR equations of motion \eqref{eom_dr}--\eqref{deltaa}.

Now, we obtain the expression $E^\text{DR}$ in the center-of-mass frame by replacing the individual positions and velocities by their expressions $\mathbf{y}_a^\text{DR}[\mathbf{y}_{12},\mathbf{v}_{12}]$ and $\dot{\mathbf{y}}_a^\text{DR}[\mathbf{y}_{12},\mathbf{v}_{12}]$. Restricting ourselves to circular orbits, the resulting CM energy depends only on $v_{12}^2=r_{12}^2\Omega^2$ and $\gamma$. Then we replace $v_{12}^2$ by its PN expansion in powers of $\gamma$ using \eqref{omegadr}, and finally replace $\gamma$ by its PN expansion \eqref{gammaxdr} in powers of $x$. We find that all poles $\propto \varepsilon^{-1}$ disappear in the process; therefore we can take the limit $\varepsilon \rightarrow 0$, and get the gauge invariant expression
\begin{align}\label{ene_dr}
    E^\text{DR} =& - \frac{m \nu c^2}{2} x  \left\{ 1 + \left( -\frac{3}{4} - \frac{\nu}{12} \right) x + \left( - \frac{27}{8} + \frac{19}{8} \nu - \frac{\nu^2}{24} \right) x^2 \right. \nonumber \\ &+ \left. \left( - \frac{675}{64} + \left[ \frac{34445}{576} - \frac{205}{96} \pi^2 \right] \nu - \frac{155}{96} \nu^2 - \frac{35}{5184} \nu^3 \right) x^3 + \mathcal{O}(c^{-8}) \right\} \,,
\end{align}
which coincides with the well-known 3PN expression of the total energy for circular orbits as given e.g. by Eq.~(192) of \cite{Bliving}.

\bibliography{}

\end{document}